\documentclass[aps,prd,reprint,superscriptaddress,floatfix]{revtex4-2}

\usepackage{babel,comment}
\usepackage{soul}
\usepackage{amsmath,amsfonts,mathrsfs,amssymb}
\usepackage{epsfig}
\usepackage{latexsym}
\usepackage{graphicx}
\usepackage{bm}
\usepackage{bbm}
\usepackage[T1]{fontenc} 
\usepackage{epsf}
\usepackage{amsthm}
\usepackage{color}
\usepackage{slashed}
\usepackage{esvect}
\usepackage{mathtools}
\usepackage{hyphenat}
\usepackage[pdfencoding=auto]{hyperref}
\usepackage[final]{microtype}
\allowdisplaybreaks
\newcommand{\RNum}[1]{\uppercase\expandafter{\romannumeral #1\relax}}
\DeclareMathOperator{\Tr}{Tr}

\begin{document}





\title[Holographic QCD EoS constrained by lattice QCD]{Holographic QCD equation of state constrained by lattice QCD: neural-ODE for probe-limit and a back-reaction test}




\author{Yutian Deng}
\thanks{Funding: start-up funding from Sun Yat-sen University (SYSU) (Y.~Deng and L.~Zhang); National Natural Science Foundation of China (NSFC) under Grant Nos.~12235016 (L.~Zhang and M.~Huang) and 12221005 (M.~Huang).}
\email[]{dengyt8@mail2.sysu.edu.cn}
\affiliation{School of Science, Shenzhen Campus of Sun Yat-sen University, No. 66, Gongchang Road, Guangming District, Shenzhen, Guangdong 518107, P.\ R.\ China}
\affiliation{Sun Yat-sen University, No. 135, Xingang Xi Road, Guangzhou, Guangdong 510275, P.\ R.\ China}

\author{Mei Huang}
\email[]{huangmei@ucas.ac.cn}
\affiliation{School of Nuclear Science and Technology, University of Chinese Academy of Sciences, Beijing, 100049, P.\ R.\ China}

\author{Lin Zhang}
\email[]{zhanglin57@mail.sysu.edu.cn}
\affiliation{School of Science, Shenzhen Campus of Sun Yat-sen University, No. 66, Gongchang Road, Guangming District, Shenzhen, Guangdong 518107, P.\ R.\ China}
\affiliation{Sun Yat-sen University, No. 135, Xingang Xi Road, Guangzhou, Guangdong 510275, P.\ R.\ China}


\begin{abstract}
  \sloppy
  We study the equation of state (EoS) of QCD matter in a bottom-up holographic setup that combines an Einstein-Maxwell-dilaton (EMD) sector with an improved Karch-Katz-Son-Stephanov (KKSS) flavor action.
  In the probe approximation, we perform an inverse reconstruction of the model functions by parameterizing them with neural networks and solving the EMD equations via a differentiable ODE solver (a neural ODE framework), calibrating the model to a $(2+1)$-flavor lattice-QCD EoS at finite temperature and finite baryon chemical potential.
  The reconstructed model functions are then parametrized and kept fixed across thermodynamic states.
  Next, viewing the EMD sector as an effective description of pure Yang--Mills theory, we fix its parameters by fitting the $\mu_B=0$ lattice pure-glue EoS using a hybrid optimization strategy.
  Finally, we go beyond the probe limit and solve the coupled EMD$+$KKSS equations with back-reaction, using the pure-glue-calibrated EMD sector as a fixed input and varying the KKSS couplings to compare with the $\mu_B=0$ two-flavor lattice EoS.
  We find a visible mismatch and a high-temperature behavior in which the back-reacted dimensionless ratios approach a nearly $\beta_1$-insensitive plateau close to the pure-glue baseline, providing a simple structural diagnostic for the present flavor-sector truncation.
\end{abstract}


\maketitle
\begin{center}
\textit{This article has been submitted to Chinese Physics C.}
\end{center}

\section{Introduction}
\label{sec_introd}

Quantum chromodynamics (QCD) is believed to be the fundamental theory of the strong interaction. In regimes of interest for heavy-ion collisions and compact-star physics, the coupling is large and controlled calculations are scarce. Lattice QCD provides high-quality results for the equation of state (EoS) at finite temperature; see, e.g., Ref.~\cite{Ding:2015ona} for a review. However, at finite baryon chemical potential $\mu_B$, the sign problem limits direct simulations, which motivates effective descriptions that can be confronted with lattice data and extrapolated to finite density. For continuum perspectives at finite temperature and density, see, e.g., Ref.~\cite{Fu:2022gou}.

Based on the AdS/CFT correspondence \cite{Maldacena:1997re}, the holographic QCD method provides a useful non-perturbative approach to explore QCD matter under extreme conditions, e.g. finite temperature and/or baryon density, magnetic and vortical fields \cite{Gubser:2008ny,DeWolfe:2010he,Jarvinen:2022doa,Yang:2015aia,Yang:2014bqa,Dudal:2017max,Dudal:2018ztm,Fang:2015ytf,Liu:2023pbt,Li:2022erd,Critelli:2017oub,Grefa:2021qvt,Arefeva:2021jpa,Arefeva:2020vae,Cai:2012xh,Chen:2019rez,He:2020fdi,Cai:2024neuralMagQCD,Cai:2022omk,Hippert:2023bel,Zhu:2025gxo,Chen:2025goz}.
In bottom-up constructions, the Einstein--Maxwell--dilaton (EMD) framework is widely used, where the dilaton field provides an effective description of the gluodynamics and the model can be calibrated to lattice thermodynamics by choosing the dilaton potential and the gauge kinetic function.
To incorporate chiral physics and flavor dynamics in addition to the gluon sector, one usually supplements the EMD background with a flavor action containing a bulk scalar field dual to $\bar q q$, as in the hard-wall/soft-wall models \cite{Erlich:2005qh,Karch:2006pv}, and their improved variants for chiral symmetry breaking and restoration \cite{Li:2013oda}.

The original soft-wall model introduced by Andreas Karch, Emanuel Katz, Dam T. Son, and Mikhail A. Stephanov \cite{Karch:2006pv}, often referred to as the Karch--Katz--Son--Stephanov (KKSS) model, provides a simple framework for light-flavor hadron spectra and chiral dynamics, and it also serves as a convenient starting point for improved soft-wall constructions.
In many applications, the KKSS sector is treated as a probe: one first solves the EMD background and then evaluates the flavor contribution on top of it. This is efficient and often sufficient for qualitative studies. A more self-consistent description requires the back-reaction of the flavor sector on the geometry. In practice, solving the coupled EMD$+$KKSS system while keeping a single action fixed across thermodynamic states, and simultaneously calibrating to lattice EoS data, is technically nontrivial. As a result, systematic back-reacted studies remain comparatively limited.
In particular, it is nontrivial to calibrate a coupled EMD$+$flavor setup to lattice thermodynamics while keeping a single five-dimensional action fixed across thermodynamic states. This motivates a controlled back-reaction study within a fixed EMD$+$KKSS action, which is one of the main focuses of this work.

In this work, we report three related studies within a single EMD$+$KKSS framework.
We first work in the probe approximation and calibrate the EMD model to the lattice EoS of $(2+1)$-flavor QCD at finite $T$ and finite $\mu_B$ by using a neural ordinary differential equation (neural ODE) approach \cite{Chen:2018wjc,kidger2022neuraldifferentialequations,Cai:2024neuralMagQCD,Zeng:2025tcz}, and subsequently fit the trained neural networks into explicit analytical expressions.
We then treat the EMD system as an effective description of pure Yang--Mills theory and fix its parameters by fitting the $\mu_B=0$ lattice pure-glue EoS via a hybrid optimization strategy; see also recent data-driven Einstein--dilaton reconstructions of pure-glue thermodynamics \cite{Chen:2025kqb}.
Finally, we go beyond the probe approximation and solve the coupled EMD$+$KKSS equations with back-reaction. With the EMD sector fixed by the pure-glue calibration, we vary the KKSS couplings and compare the resulting $\mu_B=0$ EoS with the two-flavor lattice data. A visible mismatch remains, which we view as a structural diagnostic for the present flavor-sector truncation.
Moreover, by examining the high-temperature behavior, we find that the back-reacted dimensionless ratios approach a nearly $\beta_1$-insensitive plateau close to the $\beta_1=0$ (pure-glue) baseline, which provides a simple diagnostic for the present flavor-sector truncation.

\begin{sloppypar}
This paper is organized as follows. In Sec.~\ref{sec_HQCD_backg} we introduce the EMD$+$KKSS setup and the thermodynamic dictionary. In Sec.~\ref{sec_probe_nf21} we present the probe-approximation calibration for $(2+1)$ flavors at finite $T$ and $\mu_B$ using neural ODE. In Sec.~\ref{sec_pureglue} we present the pure-glue calibration of the EMD sector at $\mu_B=0$. In Sec.~\ref{sec_back-reaction_nf2} we describe the coupled EMD$+$KKSS system with back-reaction and compare with the two-flavor $\mu_B=0$ lattice EoS. We conclude and discuss outlook in Sec.~\ref{sec_conclusion}.
\end{sloppypar}

\section{The holographic framework and the background space\hyp{}time of the holographic models}
\label{sec_HQCD_backg}

Firstly we introduce the framework of the Einstein-Maxwell-dilaton (EMD) system, which reduces to the gravity-dilaton system at zero chemical potential. The EMD system is widely used in holographic QCD models. It can be used to describe the QCD confinement-deconfinement phase diagram \cite{Cai:2012xh,Chen:2019rez}, the effect of the magnetic field on the QCD phase transition \cite{He:2020fdi,Cai:2024neuralMagQCD}, the rotating effect on the deconfinement phase transition \cite{Chen:2020ath,Sun:2023rotationSplit}, the thermodynamic properties of QCD and the location of the critical endpoint (CEP) \cite{Cai:2022omk,Hippert:2023bel,Zhu:2025gxo,Chen:2025goz}, the critical behavior near the CEP \cite{Li:2017ple,Chen:2018vty}, and the glueball spectra and corresponding EoS \cite{Zhang:2021itx,Chen:2025kqb}. Related bottom-up frameworks with dynamical flavors include the Einstein--dilaton--flavor (without a Maxwell sector) and the Einstein--Maxwell--Dilaton--flavor constructions, which have been applied to the $2{+}1$-flavor QCD phase structure \cite{Shen:2025yrn,Shen:2025zkj}.

Including the flavor sector, the total action of the $5$-dimensional holographic QCD model is
\begin{align}
   & S_{\text {total }}^{s}=S_{EMD}^{s}+S_{f}^{s},
  \label{action_total}
\end{align}
where $S_{EMD}^{s}$ is the action for the EMD system in the string frame, and $S_{f}^{s}$ is the action for the flavor part (improved KKSS type) in the string frame. Some aspects of a similar model were reviewed in Ref. \cite{Chen:2022goa}.


The EMD action in the string frame is denoted by $S_{EMD}^{s}$:
\begin{align}
  S_{EMD}^{s}=&\frac{1}{2 \kappa_{5}^{2}} \int \mathrm{d}^{5} x \sqrt{-g^{s}} e^{-2 \Phi} \Big[ R^{s}
  \nonumber\\
  &+4 {g^{s}}^{M N} \partial_{M} \Phi \partial_{N} \Phi -V^{s}(\Phi)
  \nonumber\\
  &-\frac{h(\Phi)}{4} e^{\frac{4 \Phi}{3}} {g^{s}}^{MP} {g^{s}}^{NQ} F_{MN} F_{PQ}\Big],
  \label{action_EMD_11}
\end{align}
where $s$ denotes the string frame, $\kappa_{5}^{2}=8 \pi G_5$, $G_5$ is the $5$-dimensional Newton constant. $g^s$ is the determinant of the metric in the string frame: $g^s=\det \left(g_{M N }^s\right)$. The metric tensor in the string frame can be extracted from
\begin{align}
   & d s^{2}=\frac{L^2 e^{2 A_{s}(z)}}{z^{2}}\bigg(-f(z) d t^{2}+\frac{d z^{2}}{f(z)}+d y_{1}^{2}
  \nonumber                                                                                       \\
   & \qquad \qquad \qquad \qquad +d y_{2}^{2}+d y_{3}^{2}\bigg),
  \label{metric_str}
\end{align}
where $L$ is the curvature radius of the asymptotic $AdS_5$ space\hyp{}time. For simplicity and without loss of generality, we assume $L=1$ in the following calculations.
$R^s$ is the Ricci curvature scalar in the string frame. The $5$-dimensional scalar field $\Phi(z)$ is the dilaton field that depends only on the coordinate $z$. $F_{M N}$ is the field strength of the $U(1)$ gauge field $A_{M}$:
\begin{align}
  F_{M N}=\partial_{M} A_{N}-\partial_{N} A_{M}.
  \label{field_strength_1}
\end{align}
The $5$-dimensional vector field $A_{M}$ is dual to the baryon number current. The function $h(\Phi)$ describes the gauge kinetic function and the function $V^{s}(\Phi)$ is the dilaton potential in the string frame.
In this paper, to avoid confusion with the flavor-sector potential, we denote
\begin{align}
  V^{s}(\Phi)\equiv V_{G}^{s}(\Phi),\qquad V^{E}(\Phi)\equiv V_{G}^{E}(\Phi),
  \label{VG_def_string_Einstein}
\end{align}
and reserve $V_{X}$ for the scalar potential of the KKSS sector.

\subsection{The Einstein-Maxwell-dilaton system}
\label{subsec_EMD}
\begin{sloppypar}
As discussed in Ref. \cite{Li:2011hp}, it is more convenient to choose the string frame when solving the vacuum expectation value of the loop operator, and it is more convenient to choose the Einstein frame to work out the gravity solution and to study the equation of state. So we apply the following Weyl transformation \cite{weyl1921raum,weyl1993space} to Eq. (\ref{action_EMD_11}):
\end{sloppypar}
\begin{align}
  g_{M N}^{s}=\mathrm{e}^{\frac{4}{3}\Phi}  g_{M N}^{E},
  \label{Weyl_transf}
\end{align}
where $g_{M N}^{E}$ is the metric tensor in the Einstein frame. The capital letter 'E' denotes the Einstein frame.
Then, Eq. (\ref{action_EMD_11}) becomes
\begin{alignat}{3}
   & S^E =\frac{1}{2 \kappa_{5}^{2}} & \int & \mathrm{d}^{5} x \sqrt{-g^E} \Big[ R^E-\frac{4}{3} {g^E}^{M N} \partial_{M} \Phi \partial_{N} \Phi
  \nonumber                                                                                                                                                                   \\
   &                                 &      & -V^E(\Phi)-\frac{h(\Phi)}{4} {g^{E}}^{MP} {g^{E}}^{NQ} F_{MN} F_{PQ}\Big],
  \label{action_EMD_2}
\end{alignat}
where the function $V^E=\mathrm{e}^{\frac{4}{3}\Phi} V^{s}$.

The coefficient of the kinetic term of the dilaton field $\Phi$ is $\frac{4}{3}$ in Eq. (\ref{action_EMD_2}), which is not canonical. So we define a new dilaton field $\phi$:
\begin{align}
   & \phi=\sqrt{\frac{8}{3}}\Phi.
  \label{new_dilaton}
\end{align}
Now the action Eq. (\ref{action_EMD_2}) becomes
\begin{align}
   & S^E =\int \mathrm{d}^{5} {\mathcal{L}}^E,
  \label{action_EMD_3}
\end{align}
where ${\mathcal{L}}^E$ is the Lagrangian density in the Einstein frame:
\begin{alignat}{3}
   & {\mathcal{L}}^E =\frac{1}{2 \kappa_{5}^{2}} &  & \sqrt{-g^E} \Big[ R^E-\frac{1}{2} {g^E}^{M N} \left( \partial_{M} \phi \right) \left( \partial_{N} \phi \right)
  \nonumber                                                                                                                                                                                       \\
   &                                             &  & -V_{\phi}(\phi)-\frac{h_{\phi}(\phi)}{4} {g^{E}}^{M \widetilde{M} }{g^{E}}^{N \widetilde{N} } F_{M N} F_{\widetilde{M} \widetilde{N}}\Big].
  \label{Lagrangian_EMD_1}
\end{alignat}
The function $V_\phi(\phi)=V^E(\Phi)$, and $h_{\phi}(\phi)=h(\Phi)$.

\begin{sloppypar}
For asymptotically AdS solutions, the UV behavior of $V_{\phi}(\phi)$ is constrained by the scaling dimension $\Delta_\phi$ of the operator dual to the dilaton $\phi$.
Expanding around $\phi\to 0$, one has
\begin{align}
  V_{\phi}(\phi)
  &= -\frac{12}{L^{2}}+\frac{1}{2}m_{\phi}^{2}\phi^{2}
  +\mathcal{O}(\phi^{4}),
  \nonumber\\
  m_{\phi}^{2}L^{2}
  &=\Delta_{\phi}(\Delta_{\phi}-4),
  \label{eq_delta_phi_mass_relation}
\end{align}
which is the usual AdS/CFT mass-dimension relation.
In bottom-up holographic QCD, $\Delta_\phi$ is treated as a phenomenological UV input: it controls how fast the background departs from AdS and thus encodes the nonconformality associated with the running coupling (or, equivalently, the trace anomaly), but it should not be identified with the perturbative QCD beta function.
In this work we allow different $\Delta_\phi$ in different calibration steps (e.g. $\Delta_\phi=3.6$ in the probe reconstruction of Sec.~\ref{sec_probe_nf21} and $\Delta_\phi=3.8$ in the pure-glue/back-reacted study).
These choices should be viewed as effective parameters optimized for the corresponding calibration targets and temperature windows, rather than as unique predictions derived from QCD in the UV.
A more quantitative running-coupling-based justification would require scanning $\Delta_\phi$ and refitting the model functions to the thermodynamic data, which we leave for a dedicated follow-up study.
\end{sloppypar}

According to Eqs. (\ref{metric_str}), (\ref{Weyl_transf}) and (\ref{new_dilaton}), we then derive the line element of the space\hyp{}time in the Einstein frame:
\begin{align}
   & d s^{2}=\frac{L^2 e^{2 A_{E}(z)}}{z^{2}}\bigg(-f(z) d t^{2}+\frac{d z^{2}}{f(z)}+d y_{1}^{2}
  \nonumber                                                                                       \\
   & \qquad \qquad \qquad \qquad +d y_{2}^{2}+d y_{3}^{2}\bigg),
  \label{metric_Einstein}
\end{align}
where
\begin{align}
   & A_E(z)=A_s(z)-\sqrt{\frac{1}{6}}\phi(z).
  \label{relat_of_A}
\end{align}
Using the variation method, we can derive the Einstein field equations and the equations of motion of $A_{M}$ and $\phi$ from the EMD action in Eqs.~(\ref{action_EMD_3})--(\ref{Lagrangian_EMD_1}). It is noticed that these equations are obtained by neglecting the back-reaction from the flavor sector $S_f^s$; otherwise, the field equations would involve additional source terms and become significantly more complex:
\begin{align}
   & R_{M N}^{E}-\frac{1}{2} g_{M N}^{E} R^{E}-\frac{1}{2} T_{M N}=0, \nonumber                                                                                                                                    \\
   & \nabla_{M}\left[h_{\phi}(\phi) F^{M N}\right]=0, \nonumber                                                                                                                                        \\
   & \partial_{M}\left[\sqrt{-g} \partial^{M} \phi\right]-\sqrt{-g}\left(\frac{\mathrm{d} V_{\phi}(\phi)}{\mathrm{d} \phi}+\frac{F_E^{2}}{4} \frac{\mathrm{d} h_{\phi}(\phi)}{\mathrm{d} \phi}\right)=0,
  \label{EoMs_1}
\end{align}
where $F_E^2 \equiv {g^E}^{MP} {g^E}^{NQ} F_{MN} F_{PQ}$.
where $T_{M N}$ is the energy-momentum tensor:
\begin{alignat}{3}
   & T_{M N} & = & \left(\partial_{M} \phi \right) \left(\partial_{N} \phi \right) -\frac{1}{2} g_{M N}^{E} {g^E}^{P Q}\left(\partial_{P} \phi\right) \left(\partial_{Q} \phi\right)
  \nonumber                                                                                                                                                                                                                   \\
   &         &   & -g_{M N}^{E} V_{\phi}(\phi) +h_{\phi}(\phi) \bigg({g^E}^{P Q} F_{M P} F_{N Q}
  \nonumber                                                                                                                                                                                                                   \\
   &         &   & -\frac{1}{4} g_{M N}^{E} {g^{E}}^{P \widetilde{P} }{g^{E}}^{Q \widetilde{Q} } F_{P Q} F_{\widetilde{P} \widetilde{Q}}\bigg).
  \label{energy_momentum_tensor_1}
\end{alignat}

We assume that all components of the vector field $A_{M}(z)$ vanish except the $t$-component $A_{t}(z)$. Substituting the metric in Eq.~(\ref{metric_Einstein}) into the equations of motion in Eq.~(\ref{EoMs_1}), we obtain the EoMs for the background fields:
\begin{align}
   & A_{t}^{\prime \prime}+A_{t}^{\prime}\left(-\frac{1}{z}+\frac{{h_{\phi}}^{\prime}}{h_{\phi}}+{A_{E}}^{\prime}\right)=0,
  \label{EoMs_2_1}                                                                                                                                                                         \\
   & f^{\prime \prime}+f^{\prime}\left(-\frac{3}{z}+3 {A_{E}}^{\prime}\right)-\frac{e^{-2 {A_{E}}} A_{t}^{\prime 2} z^{2} h_{\phi}}{L^2}=0,
  \label{EoMs_2_2}                                                                                                                                                                         \\
   & A_{E}^{\prime \prime}-A_{E}^{\prime}\left(-\frac{2}{z}+A_{E}^{\prime}\right)+\frac{\phi^{\prime 2}}{6}=0,
  \label{EoMs_2_4}                                                                                                                                                                         \\
   & \phi^{\prime \prime}+\phi^{\prime}\left(-\frac{3}{z}+\frac{f^{\prime}}{f}+3 A_{E}^{\prime}\right)
   \nonumber\\
   & -\frac{L^2 e^{2 {A_{E}}}}{z^{2} f} \frac{\mathrm{d} V_{\phi}(\phi)}{\mathrm{d} \phi}
   +\frac{z^{2} e^{-2 {A_{E}}} A_{t}^{\prime 2}}{2 L^2 f} \frac{\mathrm{d} h_{\phi}(\phi)}{\mathrm{d} \phi}=0,
  \label{EoMs_2_5}                                                                                                                                                                         \\
   & A_{E}^{\prime \prime} +\frac{f^{\prime \prime}}{6 f}
   +A_{E}^{\prime}\left(-\frac{6}{z}+\frac{3 f^{\prime}}{2 f}\right)
   \nonumber\\
   & \quad -\frac{1}{z}\left(-\frac{4}{z}+\frac{3 f^{\prime}}{2 f}\right)
   +3 {A_{E}}^{\prime 2}
   +\frac{L^2 e^{2 {A_{E}}} V_{\phi}}{3 z^{2} f}=0.
  \label{EoMs_2_3}                                                                                                                                                                         
\end{align}
\begin{sloppypar}
Here and in the following, a prime denotes the derivative with respect to the holographic coordinate $z$.
Only 4 equations are independent in the above 5 equations. So we choose Eq. (\ref{EoMs_2_3}) as a constraint and use it to check the solutions of the EoMs.
\end{sloppypar}

\subsection{The flavor part action in the holographic model}
\label{subsec_flavor_part}

\begin{sloppypar}
As mentioned earlier, $S_{f}^{s}$ in Eq.~(\ref{action_total}) describes the flavor part in the string frame. Here we use an improved KKSS action:
\end{sloppypar}
\begin{equation}
  \begin{aligned}
    S_{f}^{s}=&-\int \mathrm{d}^5 x \sqrt{-g^s} e^{-\Phi} \beta(\Phi)
    \\
    &\times \Tr \Bigg\{ \left| D_M X \right|^2 + V_X^s(X, \Phi, F_s^2)
    \\
    &\qquad\quad +\frac{1}{4 g_5^2} \left( F_L^2+F_R^2 \right) \Bigg\} + S_{\text{baryons}}^{s}.
  \end{aligned}
  \label{action_matter}
\end{equation}
where
\begin{align}
  D_M X=\nabla_M X-\mathrm{i} g_c A_M X
  \label{covar_der}
\end{align}
is the covariant derivative of the $5$-dimensional scalar field $X$. In this work we keep the general polynomial form
\begin{equation}
  \begin{aligned}
    V_X^s(X, \Phi, F_s^2) &= \sum_{n=1}^{2} \lambda_{2n} |X|^{2n},
  \end{aligned}
  \label{potent_X}
\end{equation}
where $\lambda_2 \equiv {M_X}_5^2$ and $F_s^2 \equiv {g^{s}}^{MP} {g^{s}}^{NQ} F_{MN} F_{PQ}$.
We remind the reader that in our convention the KKSS parameters $\lambda_2$ and $\lambda_4$ are precisely the coefficients entering Eq.~(\ref{potent_X}), controlling the quadratic and quartic terms in $V_X$.
Moreover, $\lambda_2$ is not a free parameter: it is fixed by the AdS/CFT mass-dimension relation for the bulk scalar $X$ (see the discussion below leading to ${M_X}_5^2=-3$), while $\lambda_4$ will be treated phenomenologically and tuned in the back-reacted $N_f=2$ study.
After the Weyl transformation in Eq.~(\ref{Weyl_transf}), the potential in the Einstein frame acquires an explicit dilaton dependence. In our convention,
\begin{align}
  V_X^E(X, \Phi, F_E^2)
  =e^{\frac{4}{3}\Phi}\,
  V_X^s\!\left(X, \Phi, e^{-\frac{8}{3}\Phi}F_E^2\right).
  \label{eq_VX_Einstein_from_string}
\end{align}
where $F_E^2 \equiv {g^{E}}^{MP} {g^{E}}^{NQ} F_{MN} F_{PQ}$.
The function $\beta(\Phi)$ controls the coupling strength between the flavor part and the EMD background, and we use the following smooth ansatz:
\begin{align}
  \beta(\Phi)={\beta_1}\frac{\arctan\!\left(\beta_2 \Phi-\beta_3\right)+\frac{\pi}{2}}{\pi},
  \label{beta_ansatz}
\end{align}
where $(\beta_1,\beta_2,\beta_3)$ are model parameters.
In this work, these parameters will be tuned in the comparison with the lattice EoS in the back-reacted $N_f=2$ study; in particular, we will vary $\beta_1$ to assess the sensitivity to the overall strength of the flavor back-reaction.
The original KKSS action is proposed in Ref.~\cite{Karch:2006pv} and it is modified in Refs.~\cite{Fang:2016nfj,Fang:2018vkp,Fang:2018axm,Fang:2019lmd} to describe the chiral phase transition and meson spectra.

According to the AdS/CFT dictionary \cite{Erlich:2005qh}, the bulk scalar field $X$ and the chiral gauge fields $A_{L,R}^M$ are dual to the relevant QCD operators at the ultraviolet (UV) boundary $z=0$. The bulk scalar field $X$ can be decomposed as
\begin{align}
   & X=\left(\frac{\chi (z)}{2}+S(x,z)\right) \mathrm{e}^{2 \mathrm{i} \pi(x,z)},
  \label{decomp_X}
\end{align}
where $\pi(x,z)=\pi^a(x,z) t^a$ is the pseudoscalar meson field and $S(x,z)$ is the scalar meson field. The field $\chi(z)$ is related to the vacuum expectation value (VEV):
\begin{align}
  \langle X\rangle=\frac{\chi}{2} I_{2},
  \label{VEV_X}
\end{align}
where $I_2$ is the $2\times 2$ identity matrix. For the $N_f=2$ flavor sector studied in this work, we assume the light quark masses are degenerate, which leads to the diagonal form $\chi \equiv \chi_u = \chi_d$. The $5$-dimensional bulk gauge field $A_{L,R}^M$ can be recombined into the vector field $V^M$ and the axial-vector field $A^M$:
\begin{align}
   & V^M=\frac{1}{2}(A_L^M+A_R^M),
   & A^M=\frac{1}{2}(A_L^M-A_R^M).
  \label{vec_and_axial_vec_filed}
\end{align}
The field-strength tensor for the vector field and the axial-vector field are
\begin{align}
   & F_{V}^{M N} =\frac{1}{2}\left(F_{L}^{M N}+F_{R}^{M N}\right)
  \nonumber                                                                                                                 \\
   & \qquad =\partial^{M} V^{N}-\partial^{N} V^{M}-\mathrm{i}\left[V^{M}, V^{N}\right]-\mathrm{i}\left[A^{M}, A^{N}\right],
  \\
   & F_{A}^{M N} =\frac{1}{2}\left(F_{L}^{M N}-F_{R}^{M N}\right)
  \nonumber                                                                                                                 \\
   & \qquad =\partial^{M} A^{N}-\partial^{N} A^{M}-\mathrm{i}\left[V^{M}, A^{N}\right]-\mathrm{i}\left[A^{M}, V^{N}\right].
  \label{vec_and_axial_vec_filed_strength}
\end{align}

According to mass-dimension relationship $M^2=(\Delta-p)(\Delta+p-4)$ and $\Delta=3$, $p=0$, the $5$-dimensional mass square of the bulk field $X$ is ${M_X}_5^2=-3$.

The action $S_{f}^{s}$ that describes the flavor part can be decomposed as
\begin{align}
   & S_{f}^{s}=S_{\chi}^{s}+S_{\text{mesons}}^{s}+S_{\text{baryons}}^{s},
  \label{action_matter_index2}
\end{align}
where
\begin{align}
   & S_{\chi}^{s}=- \int \mathrm{d}^5 x  \sqrt{-g^s}\,\mathrm{e}^{-\Phi}\,\beta(\Phi) \bigg\{  \frac{1}{2} \left| \partial_M \chi  -\mathrm{i} g_c A_M \chi \right|^2
  \nonumber                                                                                                                                              \\
   & \qquad\qquad +\Tr\!\left[V_X^s\!\left(\langle X\rangle,\Phi,F_s^2\right)\right] \bigg\}
  \label{action_chi}
\end{align}
After the Weyl transformation in Eq.~(\ref{Weyl_transf}), the corresponding action in the Einstein frame can be written as
\begin{align}
  S_{\chi}^{E}
  =&-\int \mathrm{d}^5 x \sqrt{-g^{E}}\,\mathrm{e}^{\Phi}\,\beta(\Phi)
  \bigg\{\frac{1}{2} \left| \partial_M \chi  -\mathrm{i} g_c A_M \chi \right|_{E}^{2}
  \nonumber\\
  &\qquad\qquad +\Tr\!\left[V_X^E\!\left(\langle X\rangle,\Phi,F_E^2\right)\right]\bigg\},
  \label{action_chi_E}
\end{align}
where $\left|\cdots\right|_{E}^{2}\equiv {g^E}^{MN}(\cdots)^\dagger(\cdots)$ and $V_X^E$ is defined in Eq.~(\ref{eq_VX_Einstein_from_string}).
Equation~(\ref{action_chi_E}) is the action for the thermodynamic VEV $\chi(z)$.
The $5$-dimensional fields dual to the meson and baryon towers are treated as perturbations around this background.
The contributions from $S_{\text{mesons}}^{s}$ and $S_{\text{baryons}}^{s}$ to the thermodynamics are expected to be subleading compared with that from $S_{\chi}^{s}$, and thus we neglect them in the thermodynamic calculation in this article.

\subsection{Coupled equations of motion with back-reaction}
\label{subsec_coupled_EoMs}

In the general case where the back-reaction of the flavor sector is considered, the fields $\{A_t, f, A_E, \phi, X\}$ are solved simultaneously as a coupled system. By varying the total action $S_{total}^E = S_{EMD}^E + S_{\chi}^E$, we can derive the complete Einstein field equations and the equations of motion for the gauge field $A_M$, the dilaton $\phi$, and the scalar field $X$. Here we focus on the thermodynamic background and only consider the contribution from the vacuum expectation value (VEV) $\chi(z)$, while the fluctuations corresponding to meson and baryon excitations are neglected in the coupled equations. In tensor form, these equations are given by:
\begin{align}
   & \frac{1}{16\pi G_5} \left[ R_{M N}^{E}-\frac{1}{2} g_{M N}^{E} R^{E} - \frac{1}{2} T_{MN}^{EMD} \right]
   \nonumber\\
   & \quad - \frac{1}{2} \beta(\Phi) e^{\Phi} T_{MN}^{\chi} = 0, 
   \label{EoMs_back_tensor_Einstein} \\
   & \partial_{M}\bigg[\sqrt{-g} \bigg(
   \frac{h_{\phi}}{16\pi G_5}
   \nonumber\\
   & \phantom{\partial_{M}\bigg[\sqrt{-g} \bigg(}
   + 4 e^{\Phi} \beta(\Phi)\,
   \frac{\partial}{\partial F_E^2}
   \Big\{\Tr\!\Big[ V_X^E(\langle X \rangle, \Phi, F_E^2) \Big]\Big\}
   \bigg) \nonumber\\
   & \phantom{\partial_{M}\bigg[} \times {g^E}^{MP} {g^E}^{NQ} F_{PQ} \bigg] \nonumber\\
   & \quad - 2 \sqrt{-g} e^{\Phi} \beta(\Phi) g_c^2 {g^E}^{NM} A_M
   \Tr\!\Big[ \langle X \rangle^\dagger \langle X \rangle \Big] = 0, 
   \label{EoMs_back_tensor_Maxwell} \\
   & \partial_{M}\left[\frac{\sqrt{-g}}{16\pi G_5} {g^E}^{MN} \partial_{N} \phi\right] \nonumber\\
   & \quad - \frac{1}{16\pi G_5} \left( \sqrt{-g} \frac{\mathrm{d} V_{\phi}}{\mathrm{d} \phi} + \sqrt{-g} \frac{F_E^{2}}{4} \frac{\mathrm{d} h_{\phi}}{\mathrm{d} \phi} \right) \nonumber\\
  & \quad - \frac{\partial}{\partial \phi} \Big\{ \sqrt{-g}\,\beta(\Phi) e^{\Phi}\,
  \Tr\!\Big[ {g^E}^{PQ} D_P \langle X \rangle^\dagger D_Q \langle X \rangle
  \nonumber\\
  & \phantom{\quad - \frac{\partial}{\partial \phi} \Big\{ \sqrt{-g}\,\beta(\Phi) e^{\Phi}\, \Tr\!\Big[}
  + V_X^E(\langle X \rangle, \Phi, F_E^2)
  \Big] \Big\}
  \nonumber\\
  & \quad = 0, 
   \label{EoMs_back_tensor_Dilaton} \\
  & D_M \Big( \sqrt{-g}\,\beta(\Phi) e^{\Phi} {g^E}^{MN} D_N \langle X \rangle \Big)
  \nonumber\\
  & \quad - \sqrt{-g}\,\beta(\Phi) e^{\Phi}\,
  \frac{\partial}{\partial \langle X \rangle^\dagger}
  \Big\{\Tr\!\Big[ V_X^E(\langle X \rangle, \Phi, F_E^2) \Big]\Big\}
  = 0,
  \label{EoMs_back_tensor_Chi}
\end{align}
where $T_{MN}^{EMD}$ is defined in Eq.~(\ref{energy_momentum_tensor_1}), and the general flavor energy-momentum tensor is defined using the trace operator:
\begin{align}
  T_{MN}^{\chi} = & \Tr \Big[ \left( D_M \langle X \rangle \right)^\dagger D_N \langle X \rangle+\left( D_N \langle X \rangle \right)^\dagger D_M \langle X \rangle \nonumber\\
  & \qquad - g_{MN}^E \Big( {g^E}^{PQ} \left( D_P \langle X \rangle \right)^\dagger D_Q \langle X \rangle \nonumber\\
  & \phantom{\qquad - g_{MN}^E \Big(} + V_X^E(\langle X \rangle, \Phi, F_E^2) \Big) \Big] \nonumber \\
  & + 4 \frac{\partial \{ \Tr [ V_X^E(\langle X \rangle, \Phi, F_E^2) ] \}}{\partial F_E^2} {g^E}^{PQ} F_{MP} F_{NQ}.
  \label{energy_momentum_tensor_chi}
\end{align}
After substituting the metric ansatz into the above tensor equations, and following the style of Eqs.~(\ref{EoMs_2_1})--(\ref{EoMs_2_5}), we obtain the following complete set of back-reacted equations of motion for $N_f=2$ flavors in the Einstein frame. To simplify the expressions, we define the effective gauge kinetic function as:
\begin{align}
  \mathcal{K}(z)
  \equiv&\, h_{\phi}
  + 64\pi G_5 e^{\Phi}\beta(\Phi)
  \frac{\partial}{\partial F_E^2}
  \Tr\!\Big[ V_X^E(\langle X \rangle, \Phi, F_E^2) \Big].
  \label{K_def}
\end{align}
The coupled equations are given by:
\begin{align}
   & A_{t}^{\prime \prime}+A_{t}^{\prime}\left(-\frac{1}{z} + A_{E}^{\prime} + \frac{\mathcal{K}'}{\mathcal{K}}\right) \nonumber\\
   & \quad - \frac{16\pi G_5 L^2 e^{2A_E} \beta(\Phi) e^\Phi g_c^2 A_t \chi^2}{z^2 f \mathcal{K}} = 0,
  \label{EoMs_back_1}                                                                                                                                                                         \\
   & f^{\prime \prime}+f^{\prime}\left(-\frac{3}{z}+3 {A_{E}}^{\prime}\right) \nonumber\\
   & \quad - \frac{16\pi G_5 \beta(\Phi) e^{\Phi} g_c^2 A_t^2 \chi^2}{f} - \frac{z^{2} e^{-2 {A_{E}}} A_{t}^{\prime 2}}{L^2} \mathcal{K} = 0,
  \label{EoMs_back_2}                                                                                                                                                                         \\
  \displaybreak[2]
   & A_{E}^{\prime \prime}-A_{E}^{\prime}\left(-\frac{2}{z}+A_{E}^{\prime}\right)+\frac{\phi^{\prime 2}}{6} \nonumber\\
   & \quad + \frac{8\pi G_5 \beta(\Phi) e^\Phi g_c^2 A_t^2 \chi^2}{3 f^2} + \frac{8\pi G_5 \beta(\Phi) e^{\Phi} \chi^{\prime 2}}{3} = 0,
  \label{EoMs_back_4}                                                                                                                                                                         \\
   & \phi^{\prime \prime}+\phi^{\prime}\left(-\frac{3}{z} + \frac{f^{\prime}}{f} + 3 A_{E}^{\prime}\right) - \frac{L^2 e^{2 {A_{E}}}}{z^{2} f} \frac{\mathrm{d} V_{\phi}}{\mathrm{d} \phi} \nonumber\\
   & \quad + \frac{z^{2} A_{t}^{\prime 2}}{2 L^2 e^{2 {A_{E}}} f} \frac{\mathrm{d} h_{\phi}}{\mathrm{d} \phi} \nonumber\\
   & \quad - 8\pi G_5 \left( \chi^{\prime 2} - \frac{g_c^2 A_t^2 \chi^2}{f^2} \right) \frac{\mathrm{d} [\beta(\Phi) e^{\Phi}]}{\mathrm{d} \phi} \nonumber\\
   & \quad - \frac{16\pi G_5 L^2 e^{2 {A_{E}}}}{z^{2} f} \frac{\partial \{ \beta(\Phi) e^{\Phi} \Tr[ V_X^E(\langle X \rangle, \Phi, F_E^2) ] \}}{\partial \phi} \nonumber\\
   & \quad = 0,
  \label{EoMs_back_5}                                                                                                                                                                         \\
  \displaybreak[2]
   & \chi'' + \chi' \left[ 3A_E' - \frac{3}{z} + \frac{f'}{f} + \Phi' \left(1 + \frac{1}{\beta(\Phi)} \frac{\mathrm{d} \beta(\Phi)}{\mathrm{d} \Phi}\right) \right] \nonumber\\
   & \quad + \frac{g_c^2 A_t^2 \chi}{f^2} - \frac{L^2 e^{2 A_E}}{z^2 f} \frac{\partial \{ \Tr[ V_X^E(\langle X \rangle, \Phi, F_E^2) ] \}}{\partial \chi} = 0,
  \label{EoMs_back_6}                                                                                                                                                                         \\
   & A_{E}^{\prime \prime} +\frac{f^{\prime \prime}}{6 f} + A_{E}^{\prime}\left(-\frac{6}{z}+\frac{3 f^{\prime}}{2 f}\right) -\frac{1}{z}\left(-\frac{4}{z}+\frac{3 f^{\prime}}{2 f}\right) \nonumber\\
   & \quad + 3 {A_{E}}^{\prime 2} - \frac{8\pi G_5 \beta(\Phi) e^\Phi g_c^2 A_t^2 \chi^2}{3 f^2} \nonumber\\
   & \quad + \frac{L^2 e^{2 {A_{E}}} \{ V_{\phi} + 16\pi G_5 \beta(\Phi) e^{\Phi} \Tr[ V_X^E(\langle X \rangle, \Phi, F_E^2) ] \}}{3 z^{2} f} \nonumber\\
   & \quad + \frac{32\pi G_5 \beta(\Phi) e^{\Phi} z^2 A_t^{\prime 2}}{3 L^2 e^{2A_E} f} \frac{\partial \{ \Tr [ V_X^E(\langle X \rangle, \Phi, F_E^2) ] \}}{\partial F_E^2} \nonumber\\
   & \quad = 0.
  \label{EoMs_back_3}
\end{align}

\begin{sloppypar}
\noindent\textbf{UV asymptotic expansions at \texorpdfstring{$z\to 0$}{z->0}.}
Near the UV boundary, the coupled equations admit fractional-power asymptotic expansions due to the non-integer scaling dimension $\Delta_{\phi}=3.8$ adopted in this work.
For convenience, we list the expansions of the five background fields $\{A_t,f,A_E,\phi,\chi\}$ around $z=0$, where the coefficients are written in terms of the canonically normalized dilaton field $\phi$ (recall $\phi=\sqrt{8/3}\,\Phi$):
The coefficients involve the EMD potential parameters $\{v_0,a_4,a_6,a_8,\dots\}$ from the pure-glue ansatz in Eq.~(\ref{eq_pureglue_VG_ansatz}) (with the best-fit set given in Eq.~(\ref{eq_pureglue_V_params})), and the gauge-kinetic parameters $\{h_0,b_1,\dots,b_{10}\}$ from Eq.~(\ref{eq_hphi_ansatz}).
\end{sloppypar}
\begin{widetext}
\begingroup\small
\begin{align}
  A_t(z)=&\,c_{A_t,0}+c_{A_t,10}\,z^{2}
  +\frac{5}{504}\,c_{A_t,10}\,c_{\phi,1}^{2}
  \left(1+\frac{42 b_{1}^{2}}{1+h_{0}}\right) z^{\frac{12}{5}}
  \nonumber\\
  &+\frac{5}{146821248}\,c_{A_t,10}\,c_{\phi,1}^{4}
  \bigg[
  -8381
  +\frac{5243616\,b_{1}^{4}}{(1+h_{0})^{2}}
  -\frac{873936}{1+h_{0}}\left(5 b_{1}^{4}-24 b_{1} b_{3}-12 b_{2}^{2} h_{0}\right)
  \nonumber\\
  &\qquad
  -\frac{218484\,b_{1}^{2}}{1+h_{0}}\left(1+300\left(a_{4}+\frac{11}{216}v_{0}\right)\right)
  -1820700\left(a_{4}+\frac{11}{216}v_{0}\right)
  \bigg] z^{\frac{14}{5}}
  +\mathcal{O}\!\left(z^{\frac{16}{5}}\right),
  \nonumber\\[0.15cm]
  f(z)=&\,1+c_{f,20}\,z^{4}
  +\frac{5}{154}\,c_{f,20}\,c_{\phi,1}^{2}\,z^{\frac{22}{5}}
  +\mathcal{O}\!\left(z^{\frac{23}{5}}\right),
  \nonumber\\[0.15cm]
  \displaybreak[2]
  A_E(z)=&\,-\frac{1}{84}\,c_{\phi,1}^{2}\,z^{\frac{2}{5}}
  +\frac{9}{64}\,c_{\phi,1}^{4}\left(\frac{239}{71442}+\frac{50}{81}\left(a_{4}+\frac{11}{216}v_{0}\right)\right)z^{\frac{4}{5}}
  \nonumber\\
  &-\frac{1}{3755372544}\,c_{\phi,1}^{6}
  \bigg[
  95401
  -254016000\left(a_{6}-\frac{1019}{77760}v_{0}\right)
  +32898600\left(a_{4}+\frac{11}{216}v_{0}\right)
  \nonumber\\
  &\qquad
  +2421090000\left(a_{4}+\frac{11}{216}v_{0}\right)^{2}
  \bigg] z^{\frac{6}{5}}
  \nonumber\\
  &+\frac{1}{36907801362432}\,c_{\phi,1}^{8}
  \bigg[
  60421987
  -335428128000\left(a_{6}-\frac{1019}{77760}v_{0}\right)
  +2300165683200\left(a_{8}+\frac{54217}{6531840}v_{0}\right)
  \nonumber\\
  &\qquad
  -308700\left(-97243+165369600\left(a_{6}-\frac{1019}{77760}v_{0}\right)\right)\left(a_{4}+\frac{11}{216}v_{0}\right)
  \nonumber\\
  &\qquad
  +4408804890000\left(a_{4}+\frac{11}{216}v_{0}\right)^{2}
  \nonumber\\
  &\qquad
  +184284639000000\left(a_{4}+\frac{11}{216}v_{0}\right)^{3}
  \bigg] z^{\frac{8}{5}}
  +\mathcal{O}\!\left(z^{2}\right),
  \nonumber\\[0.15cm]
  \phi(z)=&\,c_{\phi,1}\,z^{\frac{1}{5}}
  -\frac{1}{12096}\,c_{\phi,1}^{3}\left(198+37800\,a_{4}+1925\,v_{0}\right)z^{\frac{3}{5}}
  \nonumber\\
  &+\frac{5}{12192768}\,c_{\phi,1}^{5}
  \bigg[
  1734
  +40824000\,a_{4}^{2}
  -6531840\,a_{6}
  +7560\,a_{4}\left(79+550\,v_{0}\right)
  +116011\,v_{0}
  +105875\,v_{0}^{2}
  \bigg] z
  \nonumber\\
  &-\frac{1}{15021490176}\,c_{\phi,1}^{7}
  \bigg[
  651899
  -\frac{1579550}{27}\left(77760\,a_{6}-1019\,v_{0}\right)
  +44559270000\left(a_{4}+\frac{11}{216}v_{0}\right)^{2}
  \nonumber\\
  &\qquad
  -\frac{1225}{81}\left(-97794+201009600\,a_{6}-2634115\,v_{0}\right)\left(216\,a_{4}+11\,v_{0}\right)
  +\frac{160015625}{972}\left(216\,a_{4}+11\,v_{0}\right)^{3}
  \nonumber\\
  &\qquad
  +\frac{172480}{27}\left(6531840\,a_{8}+54217\,v_{0}\right)
  \bigg] z^{\frac{7}{5}}
  +\mathcal{O}\!\left(z^{\frac{9}{5}}\right),
  \nonumber\\[0.15cm]
  \chi(z)=&\,c_{\chi,5}\,z
  +\frac{5}{6\sqrt{6}}\,c_{\phi,1}\,c_{\chi,5}
  \left[
  21+\frac{2\beta_{2}}{(1+\beta_{3}^{2})\left(\pi-2\arctan\beta_{3}\right)}
  \right] z^{\frac{6}{5}}
  \nonumber\\
  &+\frac{5}{1344}\,c_{\phi,1}^{2}\,c_{\chi,5}
  \bigg[
  8175
  +\frac{1624\beta_{2}}{(1+\beta_{3}^{2})\left(\pi-2\arctan\beta_{3}\right)}
  +\frac{42\beta_{2}^{2}}{(1+\beta_{3}^{2})^{2}\left(\pi-2\arctan\beta_{3}\right)^{2}}
  \nonumber\\
  &\qquad
  \times\left(-1+3\pi\beta_{3}-6\beta_{3}\arctan\beta_{3}\right)
  \bigg] z^{\frac{7}{5}}
  +c_{\chi,8}\,z^{\frac{8}{5}}
  +c_{\chi,9}\,z^{\frac{9}{5}}
  +c_{\chi,10}\,z^{2}
  \nonumber\\
  &+c_{\chi,11}\,z^{\frac{11}{5}}
  +c_{\chi,12}\,z^{\frac{12}{5}}
  +c_{\chi,13}\,z^{\frac{13}{5}}
  +c_{\chi,14}\,z^{\frac{14}{5}}
  \nonumber\\
  &+\left[c_{\chi,15}+c_{\chi,15}^{(\log)}\ln z\right]z^{3}
  +\mathcal{O}\!\left(z^{\frac{16}{5}}\right).
  \label{eq_UV_asymptotic_expansions}
\end{align}
\endgroup
\end{widetext}

\begin{sloppypar}
The UV coefficients appearing in Eq.~(\ref{eq_UV_asymptotic_expansions}) have direct boundary interpretations.
For the $U(1)$ gauge field, the AdS/CFT dictionary gives the chemical potential as $\mu=A_t(z=0)$, thus
\begin{align}
  c_{A_t,0}=\mu.
  \label{eq_cAt0_is_mu}
\end{align}
The next coefficient $c_{A_t,10}$ is related to the baryon density rather than to $\mu$.
In the coupled system, the baryon density $n_{B}$ is defined from the total Lagrangian density; see Eq.~(\ref{n_B_coupled}) in Sec.~\ref{subsec_thermo_dictionary_coupled}.
Substituting the UV expansion $A_t(z)=c_{A_t,0}+c_{A_t,10}z^2+\cdots$ into Eq.~(\ref{n_B_coupled}), one can directly obtain the relation between $n_{B}$ and $c_{A_t,10}$ (see the discussion below Eq.~(\ref{n_B_coupled})).
\end{sloppypar}

\begin{sloppypar}
In our numerical calculation we fix the dimensionless coefficient $c_{\phi,1}^{\mathrm{num}}=1$ and restore physical units by multiplying any quantity with energy dimension $[X]=E^p$ by $\Lambda^p$ (see Sec.~\ref{subsec_model_input} for our convention).
In the following, we use the physical convention and simply write $c_{\phi,1}=\Lambda$ for the corresponding dimensionful UV scale, so that all UV coefficients are presented in the same physical normalization.
\end{sloppypar}

\begin{sloppypar}
For the chiral scalar, the leading UV term $\chi(z)=c_{\chi,5}z+\cdots$ plays the role of the source and is proportional to the (degenerate) $u/d$ current quark mass $m_u$ in the $N_f=2$ setup.
The coefficient of the $z^{3}$ term (together with the logarithmic piece $c_{\chi,15}^{(\log)} z^{3}\ln z$) encodes the chiral condensate $\sigma_u$.
It is noticed that in the present back-reacted setup the precise proportionality between $\{c_{\chi,15},c_{\chi,15}^{(\log)}\}$ and the condensate requires holographic renormalization, because the on-shell action contains UV divergences and the prefactor $e^{\Phi}\beta(\Phi)$ in the Einstein-frame flavor action affects the normalization of the canonical momentum.
In our numerical implementation we therefore treat $c_{\chi,5}$ as the input source parameter and use the $z^{3}$ coefficient as an indicator for the condensate; a dedicated holographic-renormalization analysis will be reported elsewhere.
\end{sloppypar}

\begin{sloppypar}
It is noticed that in Eq.~(\ref{eq_UV_asymptotic_expansions}) the truly independent UV data can be chosen as
$\{c_{A_t,0},c_{A_t,10},c_{\phi,1},c_{\chi,5},c_{\chi,15}\}$.
All other coefficients in the UV series, such as $c_{\chi,8}$--$c_{\chi,14}$ and $c_{\chi,15}^{(\log)}$, are fixed by the coupled equations order by order once the above independent data and the model parameters are specified; their explicit expressions are lengthy and we do not list them here.
\end{sloppypar}

\begin{sloppypar}
In practice, for a given horizon position $z_h$ we solve the coupled boundary-value problem by fixing $c_{A_t,0}$ (set by the chemical potential $\mu$), $c_{\chi,5}$ (set by the current quark mass $m_u$), and $c_{\phi,1}$ (taken as the UV scale $\Lambda$ in physical units), together with the standard boundary conditions $A_t(z_h)=0$, $f(z=0)=1$, and $f(z_h)=0$.
The remaining conditions are provided by the UV asymptotic AdS normalizations and the regularity conditions at the horizon, which close the system for the five second-order equations.
In our numerical implementation the coupled equations are solved using a pseudospectral (collocation) method \cite{Trefethen:2000spectral}.
After obtaining the background solution, the temperature $T$ (from $f'(z_h)$), the baryon density $n_B$ (from Eq.~(\ref{n_B_coupled})), and the chiral-condensate indicator (from the $z^3$ coefficient of $\chi$) can be extracted.
\end{sloppypar}

In the above equations, the prime $'$ denotes the derivative with respect to $z$. Only 5 equations are independent in the above 6 equations. So we choose Eq.~(\ref{EoMs_back_3}) as a constraint and use it to check the solutions of the EoMs.

Besides the second-order constraint Eq.~(\ref{EoMs_back_3}), one can also derive a simplified first-order constraint equation directly from the $zz$ component of the Einstein equations as:
\begin{align}
  & 4 \left( A_{E}^{\prime} - \frac{1}{z} \right) \left( A_{E}^{\prime} - \frac{1}{z} + \frac{f^{\prime}}{4 f} \right) - \frac{1}{6} \phi^{\prime 2} \nonumber\\
  & \quad - \frac{8\pi G_5}{3} \beta(\Phi) e^{\Phi} \left( \chi^{\prime 2} + \frac{g_c^2 A_t^2 \chi^2}{f^2} \right) \nonumber\\
  & \quad + \frac{L^2 e^{2 A_E}}{3 z^2 f} \left[ V_{\phi} + 16\pi G_5 \beta(\Phi) e^{\Phi} \Tr [ V_X^E(\langle X \rangle, \Phi, F_E^2) ] \right] \nonumber\\
  & \quad + \frac{z^2 A_t^{\prime 2}}{3 L^2 e^{2A_E} f} \bigg( \frac{1}{2} h_{\phi} \nonumber\\
    & \qquad + 64\pi G_5 \beta(\Phi) e^{\Phi} \frac{\partial \{ \Tr [ V_X^E(\langle X \rangle, \Phi, F_E^2) ] \}}{\partial F_E^2} \bigg) = 0.
  \label{EoMs_back_constraint_1st}
\end{align}
By using the equations of motion Eq.~(\ref{EoMs_back_2}) and Eq.~(\ref{EoMs_back_4}) to eliminate the second-order derivative terms $f''$ and $A_E''$ in Eq.~(\ref{EoMs_back_3}), one can show that it is completely equivalent to the first-order $zz$ constraint equation Eq.~(\ref{EoMs_back_constraint_1st}).

In our numerical study of the back-reacted $N_f=2$ theory, these coupled equations are solved by imposing consistency with the pure-glue background in the limit where the flavor contribution is small.

\subsection{Thermodynamic dictionary of the EMD system}
\label{subsec_thermo_dictionary}

We can get the black\hyp{}hole solution from the EoMs Eq. (\ref{EoMs_2_1})-(\ref{EoMs_2_5}). The temperature of the black\hyp{}hole is
\begin{align}
   & T=\left| \frac{f'(z_h)}{4\pi} \right|,
  \label{temperature}
\end{align}
where $z_h$ is the location of the horizon for the black\hyp{}hole. The entropy density of the black\hyp{}hole is
\begin{align}
   & s_{EMD}=\frac{\mathrm{e}^{3 A_E(z_h)}}{\frac{\kappa_5^2}{2\pi} z_h^3}.
  \label{entropy_density}
\end{align}
According to the AdS/CFT dictionary, the temperature and the entropy density of the QCD matter are $T$ and $s_{EMD}$, respectively.

According to the AdS/CFT dictionary, the chemical potential is
\begin{align}
   & \mu=A_t(z=0),
  \label{mu_B}
\end{align}
and the baryon number density is
\begin{align}
   & n_{EMD}= \left| \lim_{z\rightarrow 0} {\frac{\partial {{\mathcal{L}}^E}}{\partial \left( \partial_z A_t \right)}} \right|
  \nonumber                                                                                                                                                       \\
   & \quad =-\frac{1}{2 {\kappa_5}^2} \lim_{z\rightarrow 0} \left[\frac{{\mathrm{e}}^{A_E (z)}}{z} h_{\phi}(\phi) \frac{\mathrm{d}}{\mathrm{d} z} A_t (z)\right],
  \label{n_B_index_1}
\end{align}
where ${\mathcal{L}}^E$ is the Lagrangian density in the Einstein frame in Eq. (\ref{Lagrangian_EMD_1}).
From the Euler-Lagrange equation of the field $A_t (z)$:
\begin{align}
  \partial_{M} {\frac{\partial {{\mathcal{L}}^E}}{\partial \left( \partial_M A_t \right)}}-\frac{\partial {{\mathcal{L}}^E}}{\partial A_t}=0,
  \label{Euler_Lagrange_equation_of_At}
\end{align}
\begin{align}
  \partial_z \left[\frac{{\mathrm{e}}^{A_E (z)}}{z} h_{\phi}(\phi) \frac{\mathrm{d}}{\mathrm{d} z} A_t (z)\right]=0.
  \label{euqation_of_conserved_Gauss_charge}
\end{align}
Eq. (\ref{euqation_of_conserved_Gauss_charge}) is actually the EoM of $A_t$ Eq. (\ref{EoMs_2_1}). Thus, we get the conserved Gauss charge associated with the field $A_t$:
\begin{align}
  Q_G=\frac{{\mathrm{e}}^{A_E (z)}}{z} h_{\phi}(\phi) \frac{\mathrm{d}}{\mathrm{d} z} A_t (z).
  \label{conserved_Gauss_charge}
\end{align}
Then we derive
\begin{align}
   & n_{EMD}=-\frac{1}{2 {\kappa_5}^2} Q_G.
  \label{n_B_index_2}
\end{align}
Sometimes the following expression of $n_{EMD}$ which is related to the quantities at the black\hyp{}hole horizon is more convenient in the calculation:
\begin{align}
   & n_{EMD}=-\frac{1}{2 \kappa_5^2}\frac{\mathrm{e}^{A_E(z_h)}}{z_h} h_{\phi}(\phi=\phi(z_h)) {A_t}^{\prime}(z_h).
  \label{n_B_index_3}
\end{align}

The internal energy density and the free energy density are
\begin{align}
   & \epsilon_{EMD}=T \,s_{EMD}-p_{EMD}+\mu \,n_{EMD},
  \nonumber                                       \\
   & \mathcal{F}_{EMD}=-p_{EMD}=\epsilon_{EMD}-T \,s_{EMD}-\mu \,n_{EMD}.
  \label{intern_and_free_energy}
\end{align}
The differential relation of the above equations are
\begin{align}
   & \mathrm{d}\epsilon_{EMD}=T \,\mathrm{d}s_{EMD}+\mu \,\mathrm{d}n_{EMD},
  \nonumber                                                                     \\
   & \mathrm{d}\mathcal{F}_{EMD}=-\mathrm{d}p_{EMD}=-s_{EMD} \,\mathrm{d}T-n_{EMD} \,\mathrm{d}\mu.
  \label{intern_and_free_energy_diff}
\end{align}
From Eq. (\ref{intern_and_free_energy_diff}), we can calculate the pressure $P$ and the internal energy density $\epsilon$. The trace anomaly is
\begin{align}
  I_{EMD}(T,\mu) &= \epsilon_{EMD}(T,\mu)-3p_{EMD}(T,\mu)
  \nonumber\\
  &= T s_{EMD}(T,\mu) + \mu n_{EMD}(T,\mu)
  \nonumber\\
  &\quad -4p_{EMD}(T,\mu).
  \label{trace_anom}
\end{align}

The square of the speed of sound is
\begin{align}
   & {{c_s}_{EMD}}^2=\frac{\mathrm{d}{p_{EMD}}}{\mathrm{d}{\epsilon_{EMD}}}.
  \label{square_of_speef_of_sound}
\end{align}

The adiabatic index $\gamma$ is
\begin{align}
   & \gamma=\frac{\mathrm{d}\,{\ln{p_{EMD}}}}{\mathrm{d}\,{\ln{\epsilon_{EMD}}}}.
  \label{adiabatic_index_gamma}
\end{align}

Here all the thermodynamic quantities are labeled with ``EMD'', which means these thermodynamic quantities are determined by the EMD system.

\subsection{Thermodynamic dictionary of the coupled \texorpdfstring{EMD$+$KKSS}{EMD+KKSS} system}
\label{subsec_thermo_dictionary_coupled}

We now briefly comment on the thermodynamic dictionary for the coupled system $S_{\mathrm{total}}^E=S_{EMD}^E+S_{f}^E$.
The thermodynamics is still determined by the black\hyp{}hole background geometry: the temperature is given by Eq.~(\ref{temperature}), and the entropy density is given by the Bekenstein--Hawking area formula, which takes the same form as Eq.~(\ref{entropy_density}) but with the back-reacted metric function $A_E(z)$ obtained from the coupled equations.
The chemical potential is still identified as the boundary value $\mu=A_t(z=0)$, i.e. Eq.~(\ref{mu_B}).

The main difference from the pure EMD case is the baryon number density.
In the coupled system it should be defined from the total Lagrangian density,
\begin{align}
  n_{B}
  &= \left| \lim_{z\rightarrow 0} {\frac{\partial {{\mathcal{L}}_{\mathrm{total}}^E}}{\partial \left( \partial_z A_t \right)}} \right|
  \nonumber\\
  &= -\frac{1}{2 {\kappa_5}^2} \lim_{z\rightarrow 0}
  \left[\frac{{\mathrm{e}}^{A_E (z)}}{z}\,\mathcal{K}(z)\, \frac{\mathrm{d}}{\mathrm{d} z} A_t (z)\right],
  \label{n_B_coupled}
\end{align}
where ${\mathcal{L}}_{\mathrm{total}}^E$ denotes the Lagrangian density corresponding to $S_{\mathrm{total}}^E$, and $\mathcal{K}(z)$ is the effective gauge kinetic function defined in Eq.~(\ref{K_def}).
Using the UV expansion $A_t(z)=\mu+c_{A_t,10}z^2+\cdots$ together with $A_E(0)=0$ and $\beta(\Phi\to 0)\to 0$ (thus $\mathcal{K}(0)=h_{\phi}(0)=1$), one finds
\begin{align}
  n_{B}=-\frac{\mathcal{K}(0)}{8\pi G_5}\,c_{A_t,10}
  =-\frac{1}{8\pi G_5}\,c_{A_t,10},
  \label{eq_nB_cAt10_relation}
\end{align}
where we used $\kappa_5^2=8\pi G_5$.
It is noticed that the Euler--Lagrange equation for $A_t$ no longer reduces to a conservation law when the flavor sector is coupled: the Maxwell equation contains source terms (see Eq.~(\ref{EoMs_back_1}) in the general case), thus the Gauss charge $Q_G$ is not conserved and Eqs.~(\ref{euqation_of_conserved_Gauss_charge})--(\ref{n_B_index_3}) do not apply to the coupled system.

Other thermodynamic relations keep the same form, e.g. $\epsilon = Ts-P+\mu n$ and $\mathrm{d}P=s\,\mathrm{d}T+n\,\mathrm{d}\mu$, with $(s,n)$ understood as the entropy and baryon density of the coupled system.

\subsection{Model input and our working conventions}
\label{subsec_model_input}

\begin{sloppypar}
In this work, we use the same EMD$+$KKSS framework but focus on two distinct approximations/interpretations.
Firstly, in the probe approximation we solve the EMD equations of motion and interpret the background as an effective description of $(2+1)$-flavor QCD matter.
Secondly, we go beyond the probe approximation and solve the coupled EMD$+$KKSS equations with back-reaction; in this step we use the pure-glue lattice EoS to fix the EMD sector, and then interpret the back-reacted solution as the $N_f=2$ theory.
\end{sloppypar}

The main model inputs are the dilaton potential and the gauge kinetic function in the EMD sector,
$V_{\phi}(\phi)$ and $h_{\phi}(\phi)$, together with the KKSS couplings $\beta(\Phi)$ and $V_X$.
In our Mathematica code, we used a convention in which many variables carry a prefix ``var''; in this paper we always drop that prefix and use the standard symbols.

The (pure) EMD system admits an overall scaling symmetry: after setting the AdS radius to unity one first obtains thermodynamic quantities in ``bulk units'', and then introduces a single characteristic energy scale $\Lambda$ to convert them to physical units, following the convention in Ref.~\cite{Critelli:2017oub}.
Any quantity $X$ with energy dimension $[X]=E^{p}$ is mapped as
\begin{align}
  X_{\mathrm{phys}}=\Lambda^{p}\,X_{\mathrm{bulk}}.
  \label{eq_Lambda_units_map}
\end{align}
In particular, $T=\Lambda\,T_{\mathrm{bulk}}$, $\mu_B=\Lambda\,(\mu_B)_{\mathrm{bulk}}$, $s=\Lambda^{3}s_{\mathrm{bulk}}$, $n_B=\Lambda^{3}(n_B)_{\mathrm{bulk}}$, and $P,\epsilon=\Lambda^{4}(P,\epsilon)_{\mathrm{bulk}}$.
Therefore, dimensionless combinations such as $s/T^{3}$ and $P/T^{4}$ do not depend on $\Lambda$.
Equivalently, $\Lambda$ may be viewed as a conversion factor between the (dimensionless) coordinate/field normalization adopted in the bulk and the physical normalization used to report results in MeV/GeV.
This freedom is a direct consequence of the overall scale invariance of the classical EMD equations (after fixing the AdS radius $L=1$): a simultaneous rescaling of boundary coordinates and intensive thermodynamic quantities leaves all dimensionless observables invariant, while it changes the numerical values of dimensionful quantities by a common factor.
In practice, once a specific calibration prescription at $\mu=0$ is chosen (e.g. matching the lattice curve of $s(T)/T^{3}$ or fixing a reference temperature in physical units), $\Lambda$ is fixed and does not represent an additional tunable parameter.
Throughout this paper, we use $\Lambda$ to denote this overall scale (not to be confused with other uses of $\Lambda$ in unrelated contexts).
\section{Probe approximation: EMD calibrated to the lattice EoS of \texorpdfstring{$(2+1)$}{(2+1)} flavors via neural ODE}
\label{sec_probe_nf21}

\begin{sloppypar}
In this section, we work in the probe approximation: we neglect the back-reaction of the KKSS flavor sector and only solve the EMD equations of motion given in Sec.~\ref{sec_HQCD_backg}. After obtaining the black hole backgrounds, we calculate the thermodynamic quantities and compare the resulting equation of state with the lattice QCD EoS for $(2+1)$ flavors at finite temperature and finite baryon chemical potential.
\end{sloppypar}

\subsection{Neural-ODE calibration strategy}
\label{subsec_neuralODE}

\begin{sloppypar}
In this part, we calibrate the EMD model to the lattice EoS using neural ODE framework \cite{Chen:2018wjc,kidger2022neuraldifferentialequations}.
The EMD equations are treated as a differentiable ODE system, and the model functions $V_{\phi}(\phi)$ and $h_{\phi}(\phi)$ are learned from data under UV/IR constraints. The parameters are optimized by minimizing the mismatch between the holographic predictions and the lattice targets.
\end{sloppypar}

\begin{sloppypar}
Once $V_{\phi}(\phi)$ and $h_{\phi}(\phi)$ are determined, we keep them fixed and use the same five-dimensional action for all black-hole solutions at different $(T,\mu_B)$.
This keeps the thermodynamic state dependence in the solutions rather than in temperature-dependent effective potentials, and makes the subsequent back-reaction study a direct probe of the limitations of the coupled ansatz.
\end{sloppypar}

In this probe calibration, we use as reference a recently published thermodynamic table for $(2+1)$-flavor QCD at finite temperature and finite baryon chemical potential, with $\mu_Q=\mu_S=0$ and $T=35.0$--$490.0~\mathrm{MeV}$ \cite{Abuali:2025tbd}.
This table is released through a public interpolation/fitting framework that combines lattice-QCD constraints with a hadron-resonance-gas (HRG) description \cite{MUSES:4DTExS}, and provides a broad and smooth domain that is convenient for a stable inverse reconstruction of the model functions.
Our conventions for the EMD background and thermodynamic extraction follow standard choices in the recent holographic EMD literature; see, e.g., Ref.~\cite{Grefa:2021qvt}.
In our training set we include the dimensionless ratio $s/T^3$ (and thus also $p/T^4$, $\epsilon/T^4$ and $I/T^4\equiv(\epsilon-3p)/T^4$ derived from thermodynamic identities) at $\mu=0$, together with $s/T^3$ and $n_B/T^3$ at finite $\mu$.

\begin{sloppypar}
To reduce degeneracies in the inverse fit, we build the model functions by combining hard UV/IR constraints with a flexible data-driven component.
Specifically, we impose the UV behavior $V_{\phi}(\phi)=-12+\frac{1}{2}m_\phi^2\phi^2+\mathcal{O}(\phi^4)$ and $h_\phi(\phi)=1+\mathcal{O}(\phi)$ at $\phi\to 0$.
The UV mass term is fixed by the scaling dimension $\Delta_\phi$ through the AdS/CFT relation in Eq.~(\ref{eq_delta_phi_mass_relation}); see also the discussion in Sec.~\ref{sec_HQCD_backg}.
In the IR we include a term that enforces asymptotically linear glueball spectra, and set $\Delta_\phi=3.6$ (thus $\nu=4-\Delta_\phi=0.4$ and $m_\phi^2=\Delta_\phi(\Delta_\phi-4)$) together with a fixed IR coefficient $k_V=-0.1$.
The remaining parts of $V_{\phi}(\phi)$ and $h_\phi(\phi)$ are then learned from data in the neural-ODE optimization.
In practice, the flexible components are represented by fully connected neural networks with a simple $1$--$24$--$24$--$1$ architecture and SiLU activation.
\end{sloppypar}

\begin{sloppypar}
For later use, we provide an explicit functional parametrization for the reconstructed model functions.
In our convention, the Einstein-frame gluon potential and the gauge kinetic function are taken as
\begin{align}
  V_{\phi}(\phi)
  =& -12+\frac{1}{2}m_\phi^2\phi^2
  +V_{\mathrm{net}}(\phi)\,\phi^4
  \nonumber\\
  &+k_V(1+\phi^2)^{\frac{1}{4}}
  \exp\!\left(\frac{\sqrt{6}}{3}\phi\right)
  \nonumber\\
  &-\bigg[
  k_V+k_V\frac{\sqrt{6}}{3}\phi+\frac{7}{12}k_V\phi^2
  \nonumber\\
  &\qquad\quad
  +\frac{13}{36}k_V\frac{\sqrt{6}}{3}\phi^3+\frac{5}{288}k_V\phi^4
  \bigg],
  \label{eq_probe_VG_param}
\end{align}
and
\begin{align}
  h_\phi(\phi)
  =&\,k_{h_1}\exp(k_{h_2}\phi)
  \nonumber\\
  &+(1-k_{h_1})\exp\!\left[-\phi\, h_{\mathrm{net}}(\phi)\right],
  \label{eq_probe_hphi_param}
\end{align}
where $V_{\mathrm{net}}(\phi)$ and $h_{\mathrm{net}}(\phi)$ denote the flexible components represented by neural networks.
The subtraction terms in Eq.~(\ref{eq_probe_VG_param}) ensure that the UV expansion matches $V_{\phi}(\phi)=-12+\frac{1}{2}m_\phi^2\phi^2+\mathcal{O}(\phi^4)$ while keeping the desired IR asymptotics.
\end{sloppypar}

\begin{figure}[tbp]
  \centering
  \includegraphics[width=\linewidth,clip=true,keepaspectratio=true]{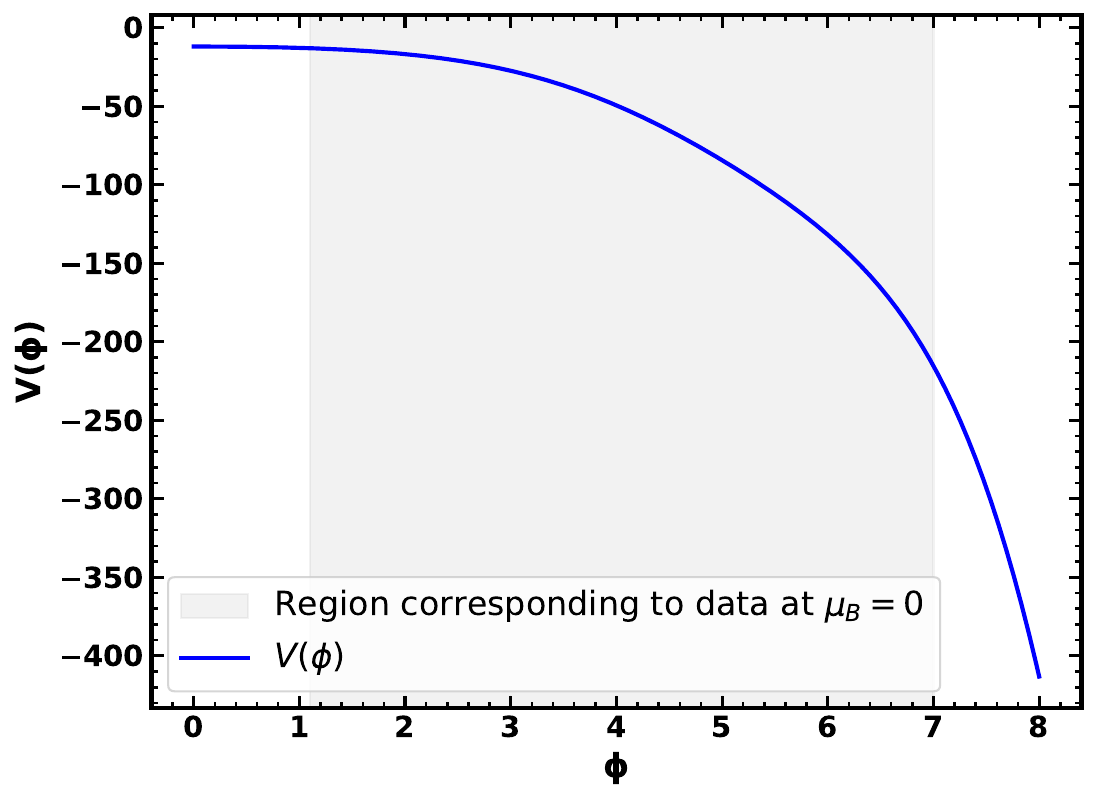}
  \par\smallskip
  \includegraphics[width=\linewidth,clip=true,keepaspectratio=true]{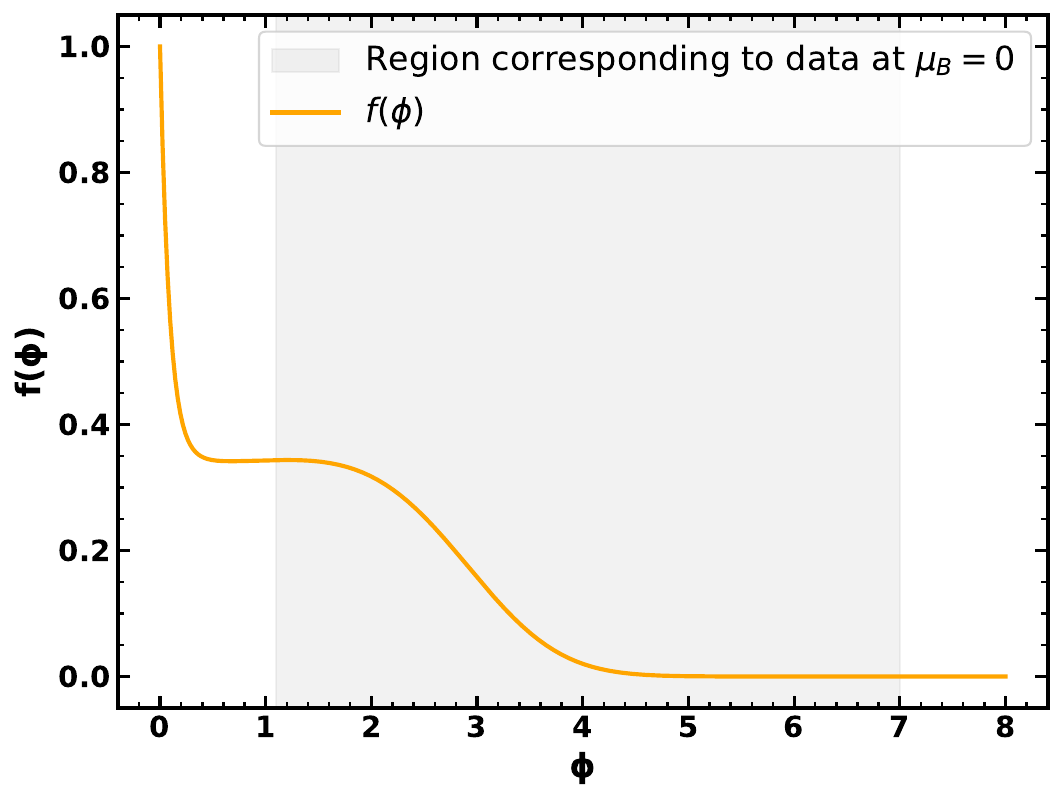}
  \caption{Probe calibration via neural ODE: reconstructed model functions $V_{\phi}(\phi)$ (upper panel) and $h_{\phi}(\phi)$ (lower panel).
  The shaded region indicates the range of $\phi$ values effectively covered by the input thermodynamic data at $\mu=0$ (in the present dataset, roughly $\phi\simeq 1.1$--$7.0$), which provides a practical measure of the domain where the reconstruction is best constrained.}
  \label{fig_probe_V_h}
\end{figure}

\begin{sloppypar}
For later reference, we fit the trained network outputs by simple polynomials (valid in the data-constrained range of $\phi$ shown in Fig.~\ref{fig_probe_V_h}):
\begin{align}
  V_{\mathrm{net}}(\phi)
  =&-1.558082\times 10^{-7}\phi^8
  \nonumber\\
  &+4.058989\times 10^{-6}\phi^7
  \nonumber\\
  &-2.014347\times 10^{-5}\phi^6
  \nonumber\\
  &-3.300564\times 10^{-4}\phi^5
  \nonumber\\
  &+4.517985\times 10^{-3}\phi^4
  \nonumber\\
  &-2.073986\times 10^{-2}\phi^3
  \nonumber\\
  &+3.797004\times 10^{-2}\phi^2
  \nonumber\\
  &+6.484887\times 10^{-4}\phi
  \nonumber\\
  &-1.702365\times 10^{-1},
  \label{eq_probe_Vnet_poly}
\end{align}
\begin{align}
  h_{\mathrm{net}}(\phi)
  =&\phantom{-}7.523421\times 10^{-7}\phi^8
  \nonumber\\
  &-3.743422\times 10^{-5}\phi^7
  \nonumber\\
  &+\phantom{-}7.309245\times 10^{-4}\phi^6
  \nonumber\\
  &-6.995335\times 10^{-3}\phi^5
  \nonumber\\
  &+\phantom{-}3.262371\times 10^{-2}\phi^4
  \nonumber\\
  &-6.210014\times 10^{-2}\phi^3
  \nonumber\\
  &+\phantom{-}9.327736\times 10^{-2}\phi^2
  \nonumber\\
  &-1.261251\times 10^{-1}\phi
  \nonumber\\
  &+\phantom{-}8.025826\times 10^{-2}.
  \label{eq_probe_fnet_poly}
\end{align}
\end{sloppypar}

\begin{sloppypar}
The best-fit constants in Eqs.~(\ref{eq_probe_VG_param})--(\ref{eq_probe_hphi_param}) are
\begin{alignat}{2}
  \Delta_\phi &= 3.6, \qquad & \nu &= 4-\Delta_\phi=0.4, \nonumber\\
  m_\phi^2 &= \Delta_\phi(\Delta_\phi-4)=-1.44, \qquad & k_V &= -0.1, \nonumber\\
  k_{h_1} &= 0.652974, \qquad & k_{h_2} &= -11.292836, \nonumber\\
  \Lambda &= 1130.97~\mathrm{MeV}, \qquad & \kappa_5^2 &= 10.39.
  \label{eq_probe_bestfit_constants}
\end{alignat}
In this probe calibration we take $\Delta_\phi=3.6$, while in the pure-glue/back-reacted steps below we adopt a different value (e.g. $\Delta_\phi=3.8$) tailored to that calibration; we keep the same notation to match the corresponding numerical setups.
\end{sloppypar}

\begin{sloppypar}
The training is performed in two stages.
First, we fix $\mu=0$ and optimize $V_{\phi}(\phi)$ by matching $s/T^3$.
Second, we freeze the trained $V_{\phi}(\phi)$ and optimize $h_\phi(\phi)$ by matching both $s/T^3$ and $n_B/T^3$ at finite $\mu$.
For a data sample at $(T_i,\mu_i)$ with uncertainty $e_i$, we define the loss functions as
\begin{align}
  L_{s/T^3}
  &=\frac{1}{N}\sum_i
  \left[
    \frac{\left(s/T^3\right)_{m_i}-\left(s/T^3\right)_{d_i}}{\left(s/T^3\right)_{e_i}}
  \right]^2,
  \label{eq_loss_sT3}
  \\
  L_{n_B/T^3}
  &=\frac{1}{N}\sum_i
  \left[
    \Big(\ln\!\left(n_B/T^3\right)_{m_i}-\ln\!\left(n_B/T^3\right)_{d_i}\Big)
  \right.
  \nonumber\\
  &\qquad\left.
    \times
    \frac{\left(n_B/T^3\right)_{d_i}}{\left(n_B/T^3\right)_{e_i}}
  \right]^2,
  \label{eq_loss_nBT3}
\end{align}
where the subscripts $m_i$, $d_i$, and $e_i$ denote the model prediction, the reference-data value, and the data uncertainty, respectively.
The logarithmic form in Eq.~(\ref{eq_loss_nBT3}) balances the wide dynamic range of $n_B/T^3$.
We propagate gradients through the differentiable EMD solver using an adjoint sensitivity method \cite{Chen:2018wjc,kidger2022neuraldifferentialequations} to efficiently optimize the network parameters.
\end{sloppypar}

\begin{figure}[tbp]
  \centering
  \includegraphics[width=0.90\linewidth,clip=true,keepaspectratio=true]{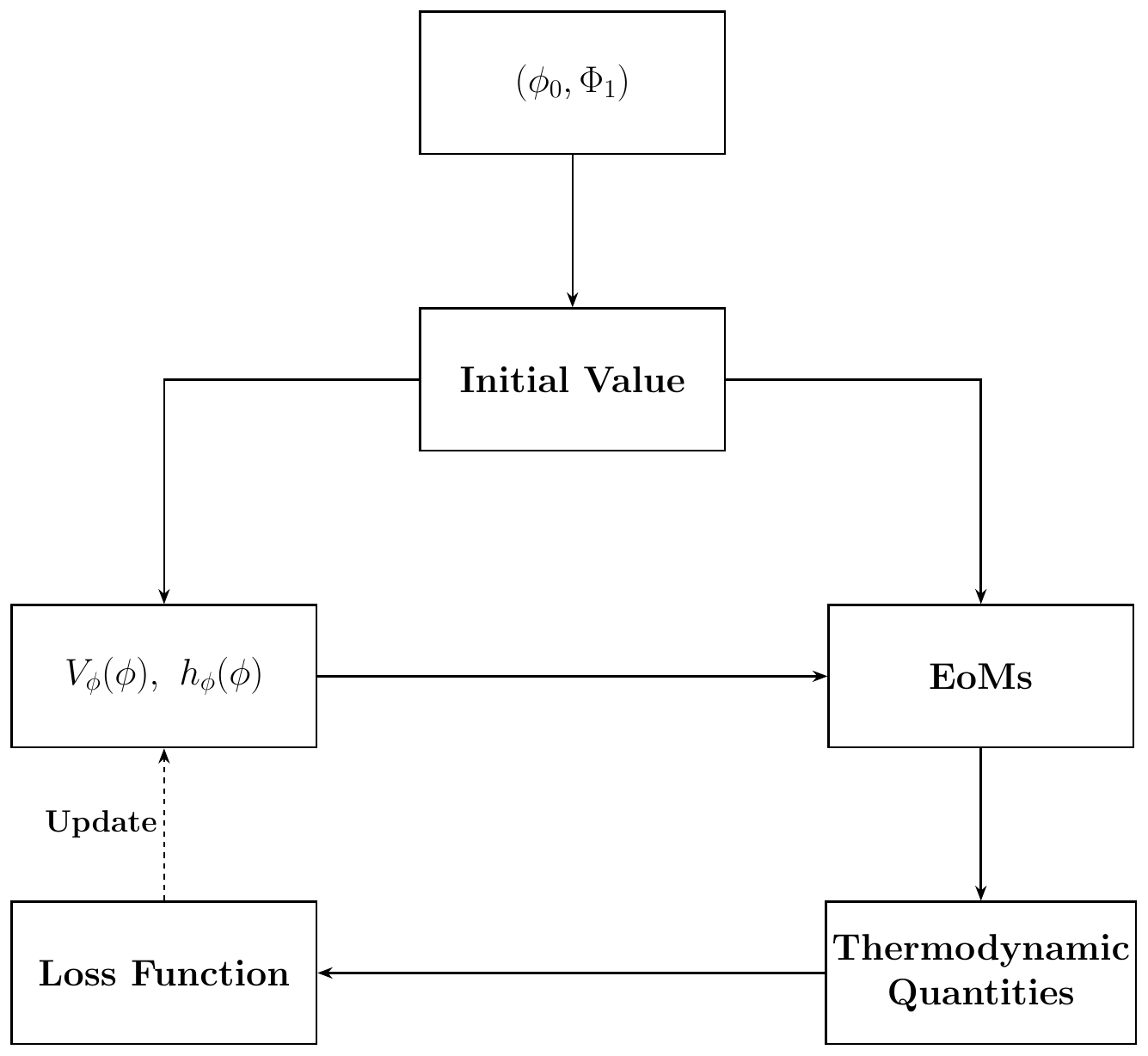}
  \caption{Schematic illustration of the neural-ODE calibration workflow.
  For a given set of model-function parameters, the EMD equations are solved to obtain thermodynamic observables, which are compared with the reference table to construct the loss, and the parameters are updated by gradient-based optimization.}
  \label{fig_training_framework}
\end{figure}

\subsection{EoS results and comparison with lattice}
\label{subsec_probe_results}

\begin{figure*}[p]
  \centering
  \includegraphics[width=0.49\textwidth,clip=true,keepaspectratio=true]{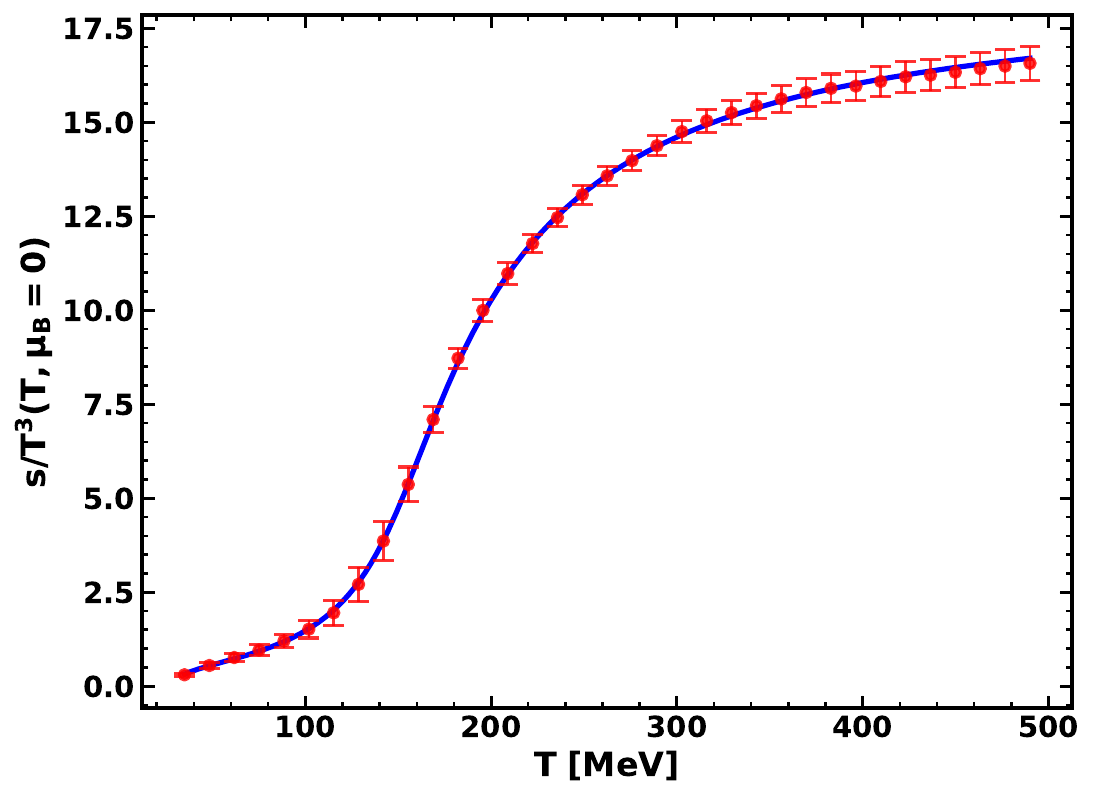}
  \hfill
  \includegraphics[width=0.49\textwidth,clip=true,keepaspectratio=true]{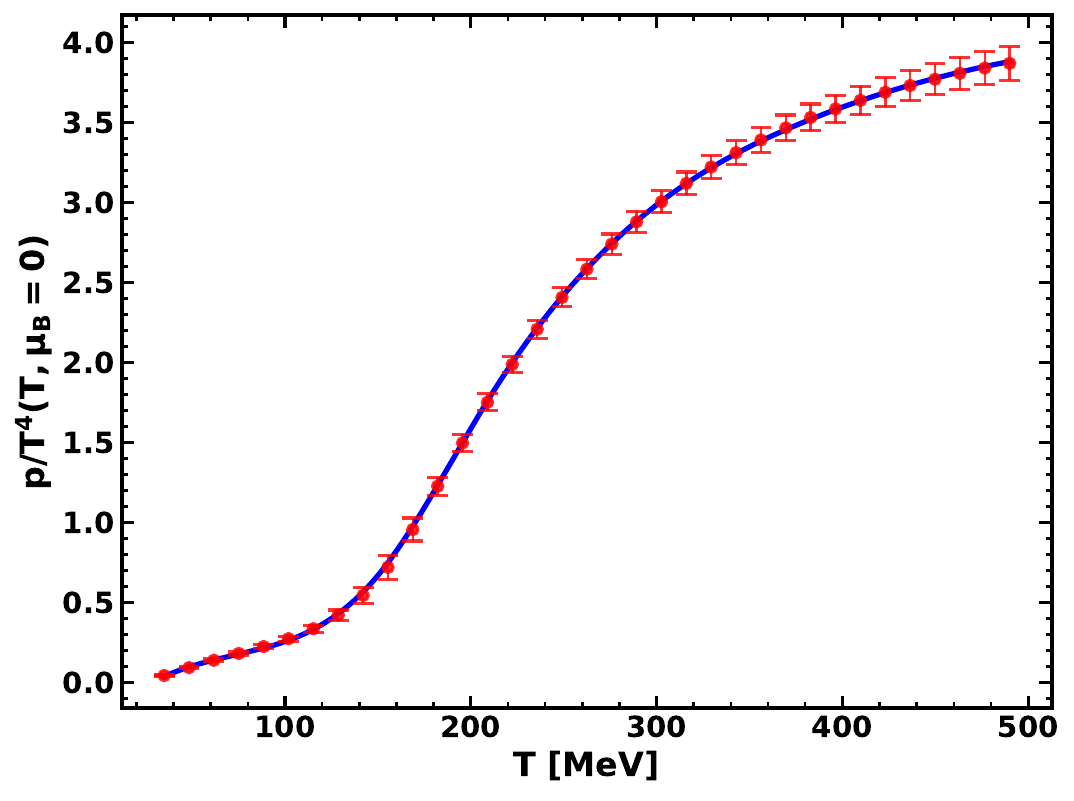}
  \vspace{0.08cm}\\
  \includegraphics[width=0.49\textwidth,clip=true,keepaspectratio=true]{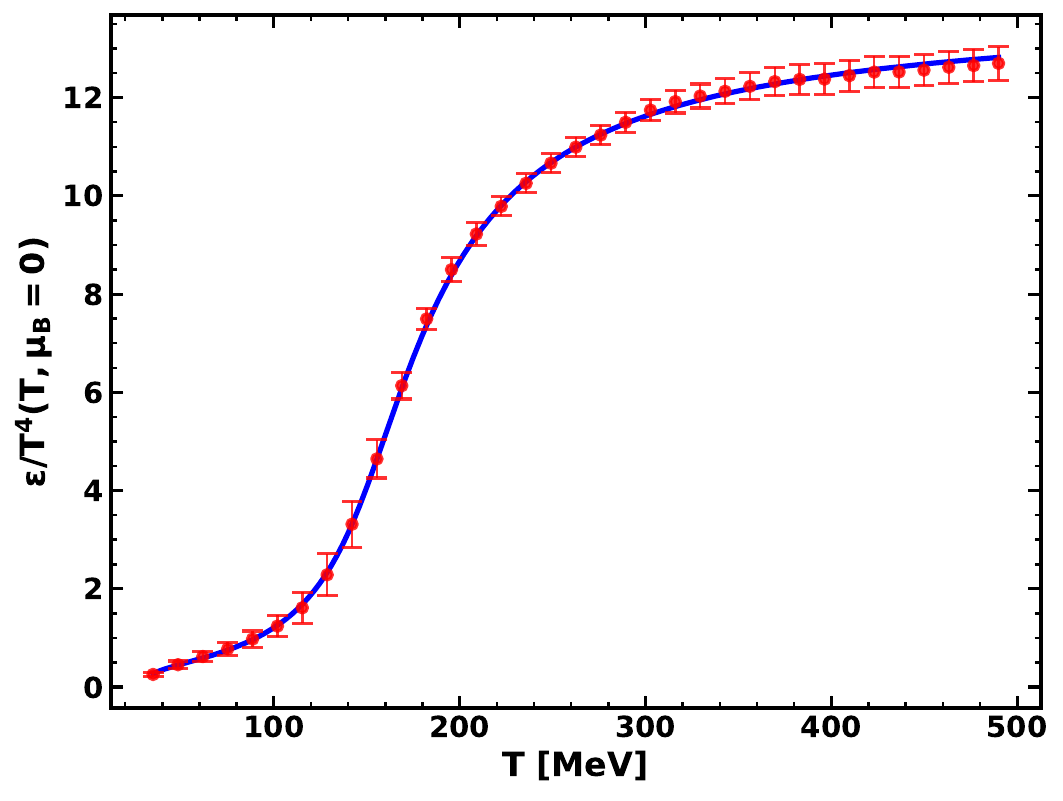}
  \hfill
  \includegraphics[width=0.49\textwidth,clip=true,keepaspectratio=true]{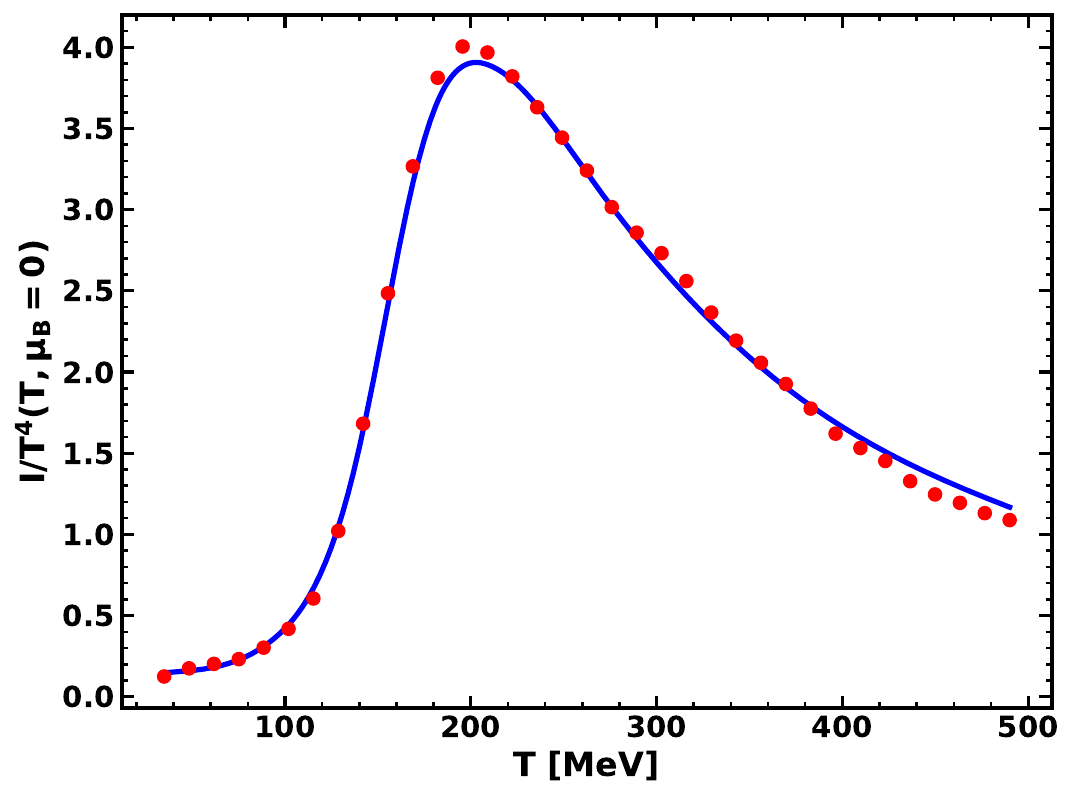}
  \caption{Probe approximation, $(2+1)$ flavors at $\mu=0$: comparison between the holographic EMD results (solid lines) and the reference thermodynamic table \cite{Abuali:2025tbd} (scatter points).
  Upper left panel: $s/T^3$; upper right panel: $p/T^4$; lower left panel: $\epsilon/T^4$; lower right panel: $I/T^4=(\epsilon-3p)/T^4$.
  The reference points for $I/T^4$ are constructed from the combination $\epsilon-3p$ and thus do not represent an independently tabulated observable.
  The horizontal axis is $T$ in units of MeV.}
  \label{fig_probe_mu0}
\end{figure*}

\begin{figure*}[p]
  \centering
  \includegraphics[width=0.49\textwidth,clip=true,keepaspectratio=true]{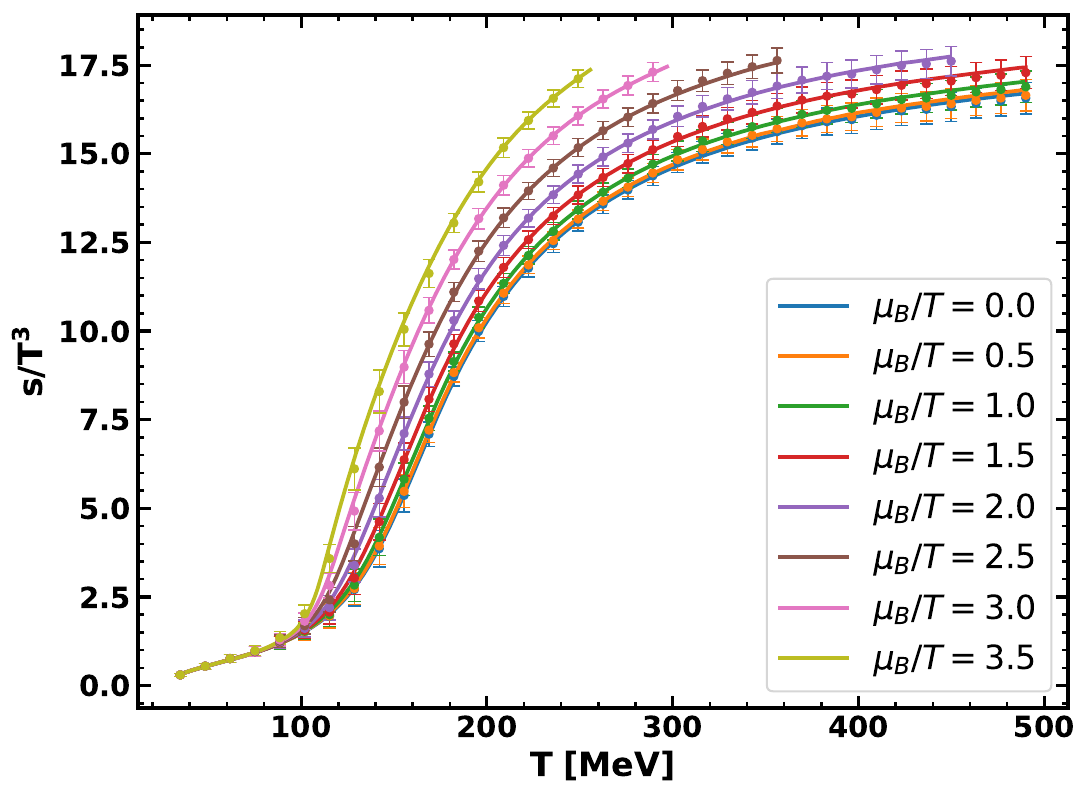}
  \hfill
  \includegraphics[width=0.49\textwidth,clip=true,keepaspectratio=true]{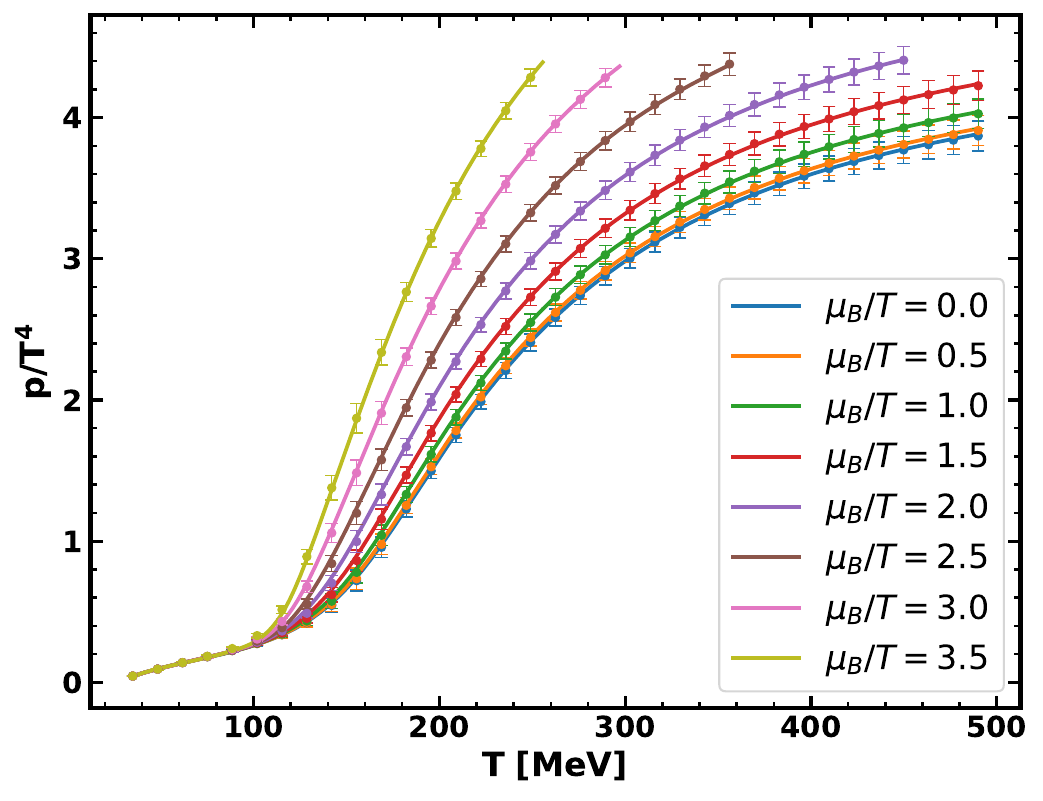}
  \vspace{0.08cm}\\
  \includegraphics[width=0.49\textwidth,clip=true,keepaspectratio=true]{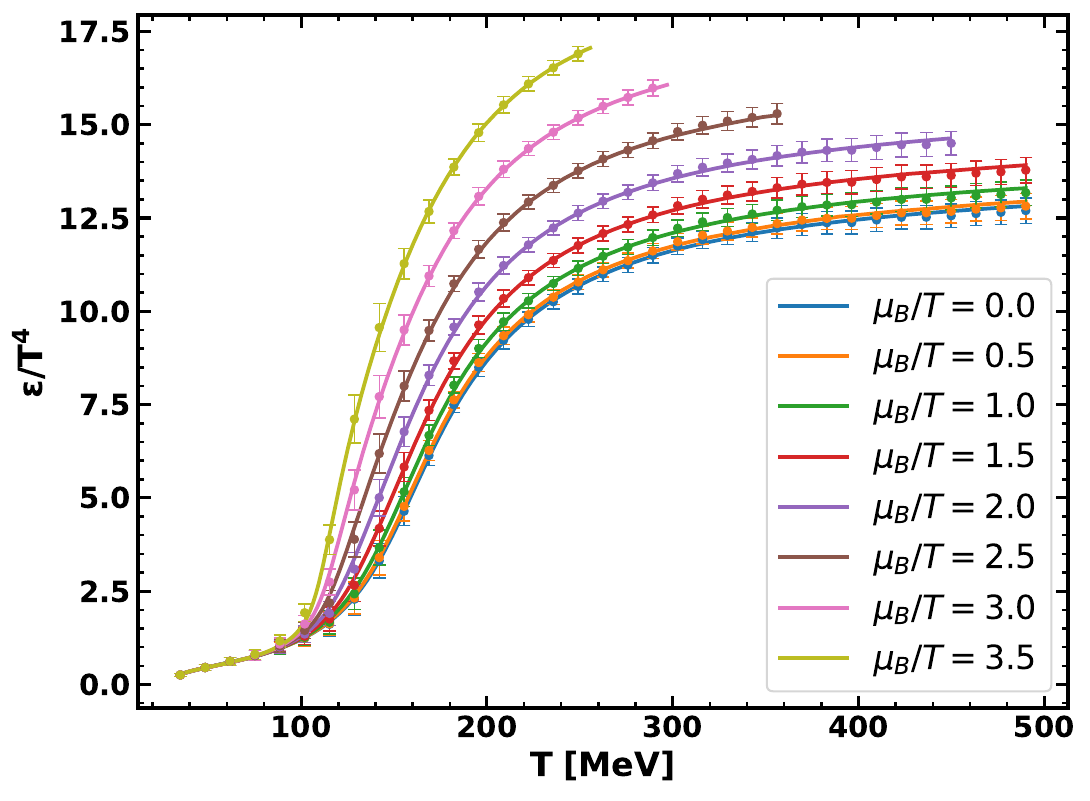}
  \hfill
  \includegraphics[width=0.49\textwidth,clip=true,keepaspectratio=true]{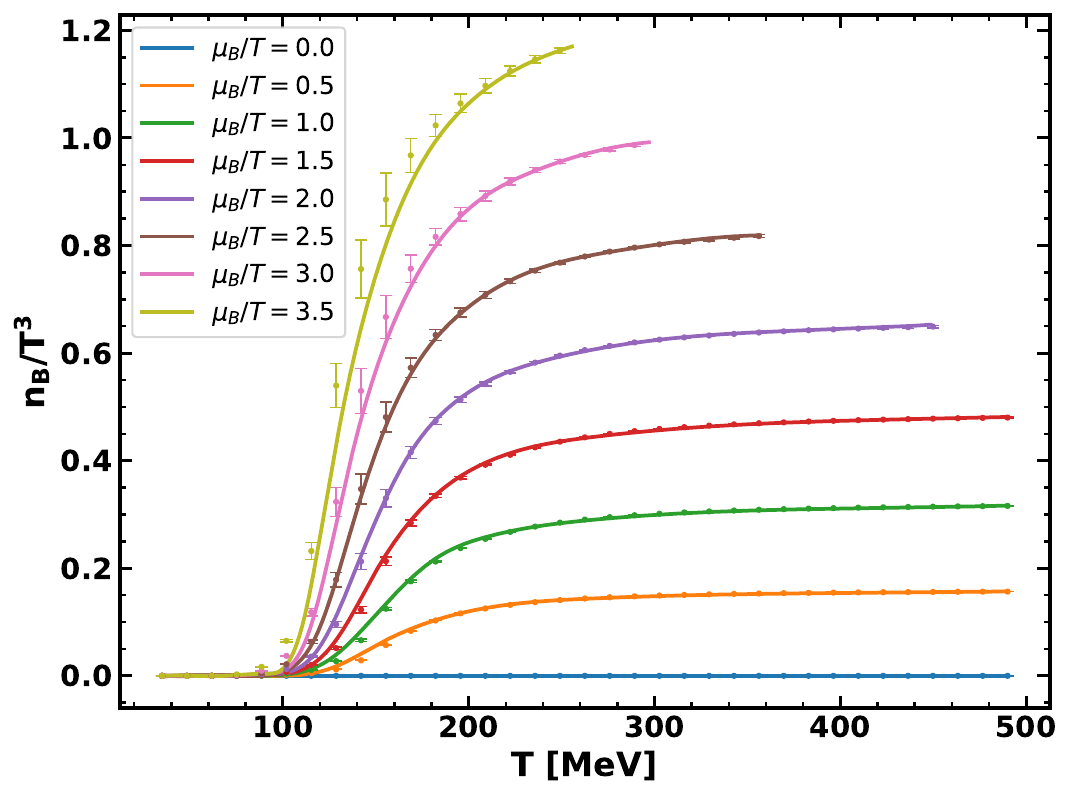}
  \caption{Probe approximation, $(2+1)$ flavors in fixed $\mu/T$ slices (temperature window $T=35$--$490~\mathrm{MeV}$): comparisons between the holographic EMD results (solid lines) and the reference thermodynamic table \cite{Abuali:2025tbd} (scatter points).
  Upper left panel: $s/T^3$; upper right panel: $p/T^4$; lower left panel: $\epsilon/T^4$; lower right panel: $n_B/T^3$.}
  \label{fig_probe_muT_slices}
\end{figure*}

\begin{sloppypar}
After training, we obtain an excellent agreement with the reference thermodynamic table for $(2+1)$-flavor QCD in a wide range of temperature and chemical potential.
In Fig.~\ref{fig_probe_mu0} we show representative comparisons at $\mu=0$, while in Fig.~\ref{fig_probe_muT_slices} we show the comparisons in fixed $\mu/T$ slices in the temperature window $T=35$--$490~\mathrm{MeV}$.
\end{sloppypar}

\begin{figure}[tbp]
  \centering
  \includegraphics[width=\linewidth,clip=true,keepaspectratio=true]{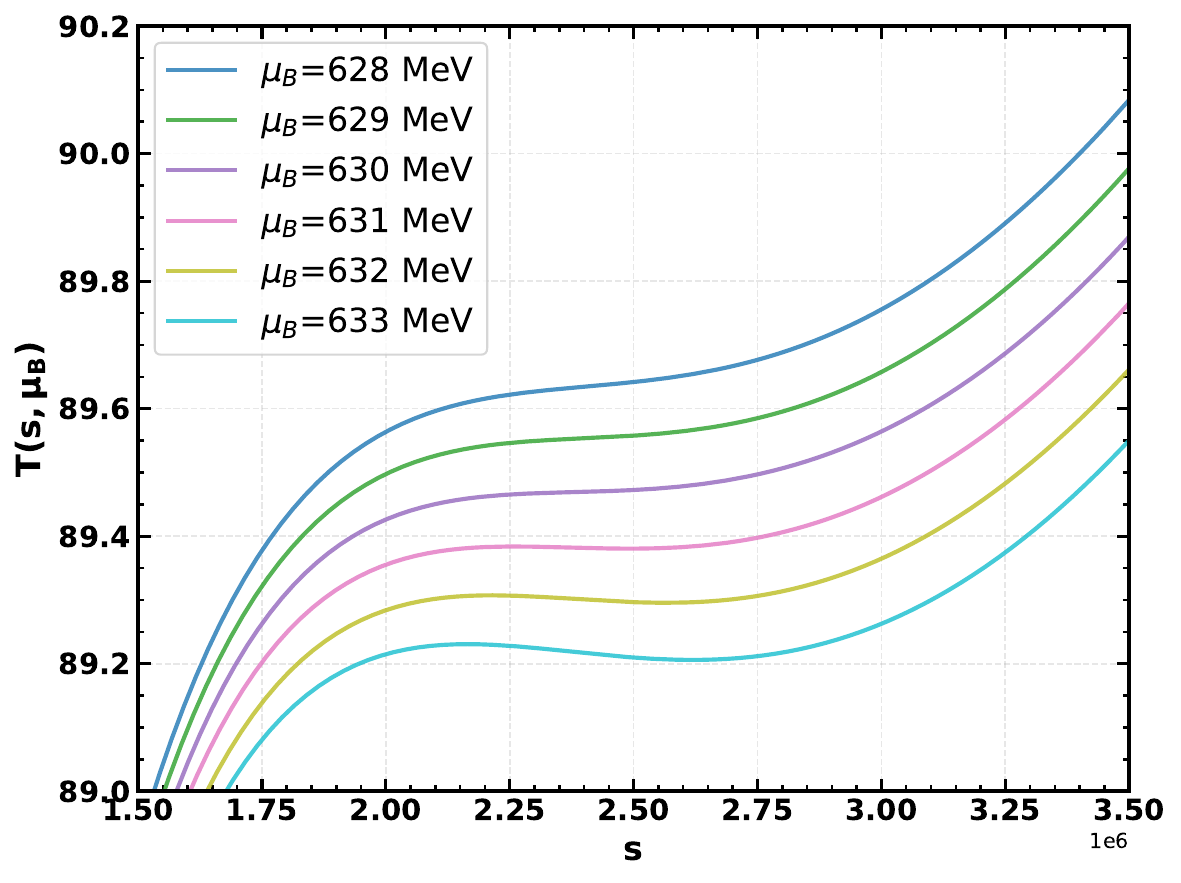}
  \par\smallskip
  \includegraphics[width=\linewidth,clip=true,keepaspectratio=true]{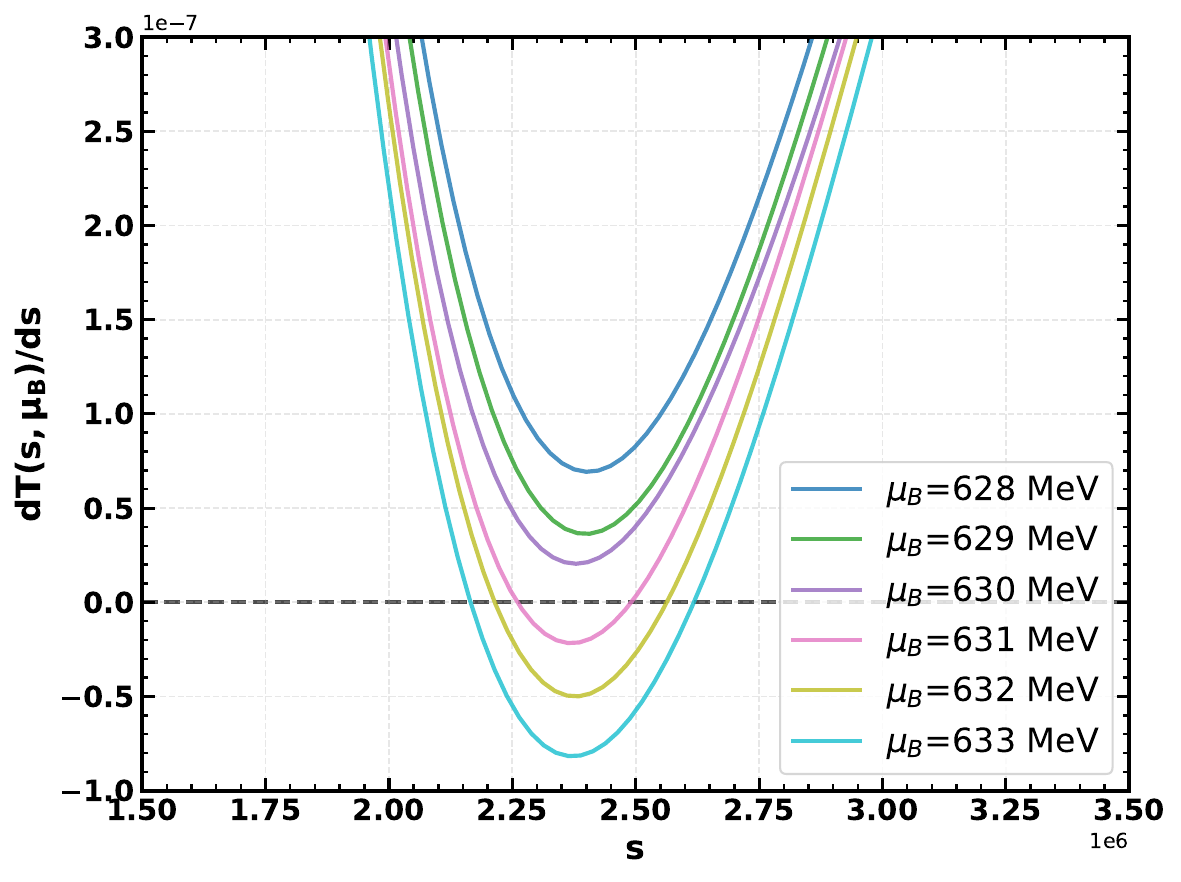}
  \caption{Determination of the QCD critical endpoint (CEP) in the probe-approximation model.
  Upper panel: $T(s,\mu_B)$ curve with the CEP marked by a red dot.
  Lower panel: $\mathrm{d}T/\mathrm{d}s$ curve, where the extremum (black star) corresponds to the CEP.}
  \label{fig_probe_CEP}
\end{figure}

\begin{sloppypar}
As an additional application of the reconstructed probe-approximation model, one may explore the thermodynamics at finite $\mu$ and infer the possible location of the critical endpoint (CEP).
Following the standard procedure, we inspect the behavior of $T(s,\mu_B)$ and identify the CEP from the extremum of $\mathrm{d}T/\mathrm{d}s$ along fixed-$\mu_B$ curves.
An illustration of this determination is shown in Fig.~\ref{fig_probe_CEP}.
With this procedure, the CEP in the present reconstructed probe model is located at approximately $(T^{\mathrm{CEP}},\mu_B^{\mathrm{CEP}})\approx (89.4\pm0.2,\ 630.5\pm1.0)\,\mathrm{MeV}$.
\end{sloppypar}

\section{Pure-glue calibration of the EMD sector at \texorpdfstring{$\mu=0$}{mu=0}}
\label{sec_pureglue}

We now consider the EMD system as a dual description of the pure Yang--Mills theory. In this step we set $\mu_B=0$, thus $A_t(z)=0$, and the Maxwell sector decouples. Therefore the gauge kinetic function $h_{\phi}(\phi)$ does not enter the thermodynamics at $\mu_B=0$, and the pure-glue calibration mainly constrains the dilaton potential $V_{\phi}(\phi)$ and the overall normalizations.

\subsection{Ansatz for \texorpdfstring{$V_{\phi}(\phi)$}{Vphi(phi)} and fitted parameters}
\label{subsec_pureglue_ansatz}

We take the following analytic ansatz in $d=4$:
\begin{align}
  V_{\phi}(\phi) =&-d(d-1)
  \nonumber\\
  &+\left[\frac{\Delta_{\phi}(\Delta_{\phi}-d)}{2}
  -\frac{v_0}{2(d-1)}\right]\phi^2
  \nonumber\\
  &+a_4 \phi^4 +a_6 \phi^6 +a_8 \phi^8 +a_{10} \phi^{10}
  \nonumber\\
  &+v_0(1+\phi^2)^{\frac{1}{4}} {\left[\sinh\!\left(\frac{\phi}{\sqrt{2(d-1)}}\right)\right]}^2,
  \label{eq_pureglue_VG_ansatz}
\end{align}
which is UV consistent by construction.

Indeed, expanding Eq.~(\ref{eq_pureglue_VG_ansatz}) around $\phi=0$ one finds
\begin{align}
  &v_0(1+\phi^2)^{\frac{1}{4}} {\left[\sinh\!\left(\frac{\phi}{\sqrt{2(d-1)}}\right)\right]}^2
  \nonumber\\
  &\quad = \frac{v_0}{2(d-1)}\,\phi^2+\mathcal{O}(\phi^4),
\end{align}
which cancels the explicit $-\frac{v_0}{2(d-1)}\phi^2$ term in the second line.
Therefore the quadratic coefficient is fixed to be $\frac{\Delta_{\phi}(\Delta_{\phi}-d)}{2}$, in agreement with the AdS/CFT mass-dimension relation. In particular, for $d=4$ one has $m_\phi^2L^2=\Delta_\phi(\Delta_\phi-4)$, see Eq.~(\ref{eq_delta_phi_mass_relation}); in our numerical implementation we set $L=1$ as discussed around Eq.~(\ref{metric_str}).
This structure ensures that the background is asymptotically AdS and that the UV scaling dimension $\Delta_{\phi}$ of the operator dual to $\phi$ is well-defined, while still allowing $v_0$ to control the nonconformal/IR behavior without spoiling the UV data.
In the present fit we take $\Delta_{\phi}=3.8$.

The best-fit parameters obtained by a hybrid optimization (Bayesian optimization combined with local refinement) are
\begin{align}
  \Lambda &= 1.19785~\mathrm{GeV},
  \nonumber\\
  G_5 &= 1.16750,
  \label{eq_pureglue_norm_params}
\end{align}
and
\begin{align}
  v_0&=-\frac{6}{5},
  \nonumber\\
  a_4&=-7.74146 \times 10^{-2},
  \nonumber\\
  a_6&=-3.48325 \times 10^{-4},
  \nonumber\\
  a_8&=-1.36440 \times 10^{-5},
  \nonumber\\
  a_{10}&=-2.89020\times 10^{-8}.
  \label{eq_pureglue_V_params}
\end{align}

Here $\Lambda$ is the overall energy scale of the EMD system introduced to convert bulk units to physical units; see Eq.~(\ref{eq_Lambda_units_map}) and Ref.~\cite{Critelli:2017oub}.

\subsection{A convenient ansatz for \texorpdfstring{$h_{\phi}(\phi)$}{h(phi)}}
\label{subsec_hphi_ansatz}

For use at finite $\mu$, we also record the functional form of the gauge kinetic function $h_{\phi}(\phi)$:
\begin{align}
  h_{\phi}(\phi)=& \frac{1}{1+h_0}\,\mathrm{sech}\bigl( b_1\phi+b_3\phi^3+b_5\phi^5
  \nonumber\\
  &\quad +b_7\phi^7+b_9\phi^9 \bigr)
  \nonumber\\
  &+\frac{h_0}{1+h_0}\,\mathrm{sech}\bigl( b_2\phi^2+b_4\phi^4+b_6\phi^6
  \nonumber\\
  &\quad +b_8\phi^8+b_{10}\phi^{10} \bigr),
  \label{eq_hphi_ansatz}
\end{align}
where the set of parameters $\{h_0, b_1, \dots, b_9, b_2, \dots, b_{10}\}$ are to be fixed.

It is noticed again that at $\mu=0$ the Maxwell sector decouples and $h_{\phi}(\phi)$ does not enter the thermodynamics.

\subsection{Pure-glue EoS at \texorpdfstring{$\mu=0$}{mu=0}}
\label{subsec_pureglue_EoS}

\begin{sloppypar}
In practice, the pure-glue calibration amounts to minimizing a chi-square objective constructed from the lattice entropy density $s/T^3$ at $\mu=0$.
Concretely, for each lattice temperature point $T_i$ (with uncertainty $\sigma_i$), we define
\begin{align}
  \chi^2=\sum_{i\ge 6}\left[\frac{
  \left.\left(\frac{s}{T^3}\right)_{\mathrm{model}}\right|_{T=T_i}
  -\left.\left(\frac{s}{T^3}\right)_{\mathrm{lat}}\right|_{T=T_i}
  }{\sigma_i}\right]^2,
\end{align}
and determine the best-fit model parameters by minimizing $\chi^2$.
In practice, the sum starts from the 6th lattice temperature point (excluding the five points below $T_c$), because their absolute uncertainties are very small and would otherwise over-weight the near-transition region in the chi-square objective.
Since each function evaluation requires solving the background and extracting thermodynamic quantities, we adopt a hybrid Bayesian-optimization strategy based on a Gaussian-process surrogate model \cite{Jones:1998ego,Shahriari:7352306,Snoek:2012pbo}, combined with local refinement.
Starting from an initial Latin-hypercube sampling in the parameter box, we iteratively fit the surrogate to the accumulated samples, propose new candidates by alternating a local acquisition-based search around the current best point (with an adaptive search radius) and occasional global exploration, and evaluate the expensive objective in parallel; failed evaluations are discarded and successful samples are cached to enable restartable runs.
\end{sloppypar}

\begin{figure*}[!t]
  \centering
  \includegraphics[width=0.48\linewidth,height=0.30\linewidth,clip=true,keepaspectratio=true]{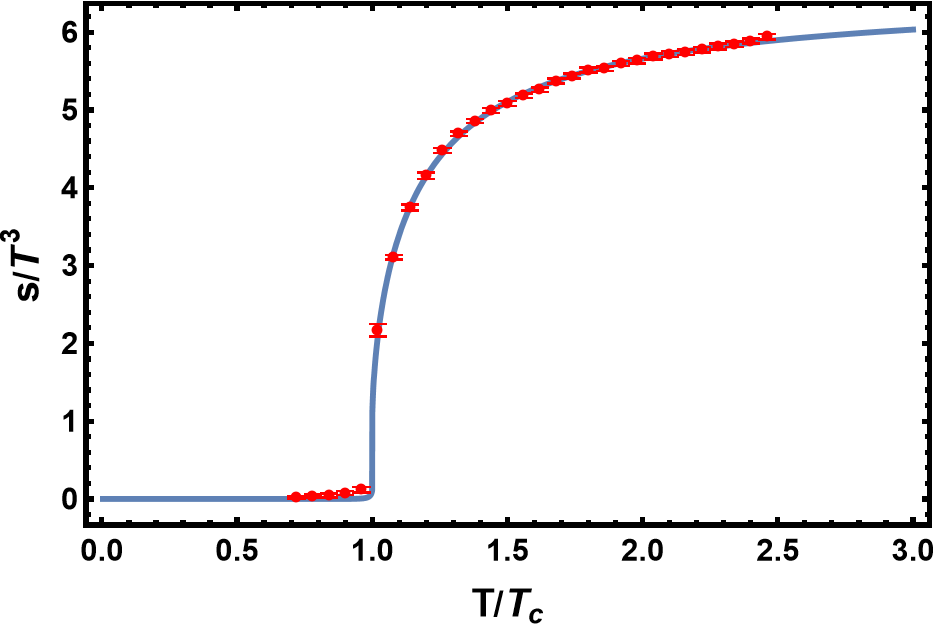}
  \hfill
  \includegraphics[width=0.48\linewidth,height=0.30\linewidth,clip=true,keepaspectratio=true]{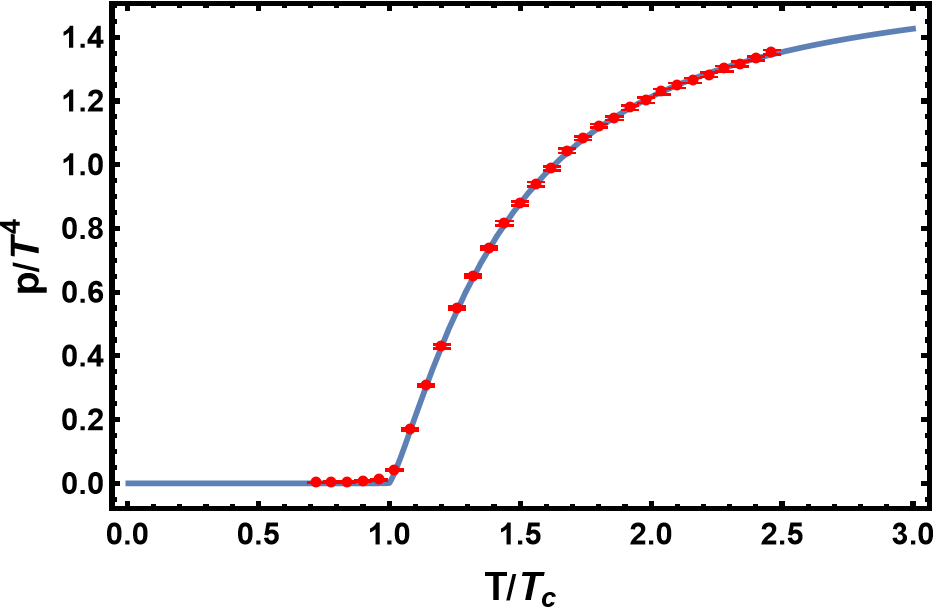}
  \vspace{0.06cm}\\
  \includegraphics[width=0.48\linewidth,height=0.30\linewidth,clip=true,keepaspectratio=true]{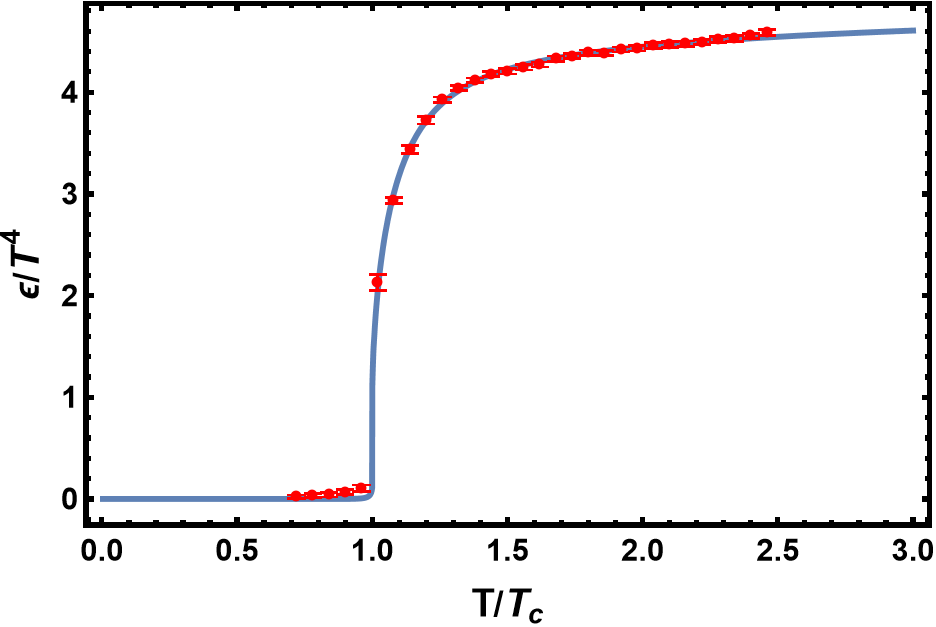}
  \hfill
  \includegraphics[width=0.48\linewidth,height=0.30\linewidth,clip=true,keepaspectratio=true]{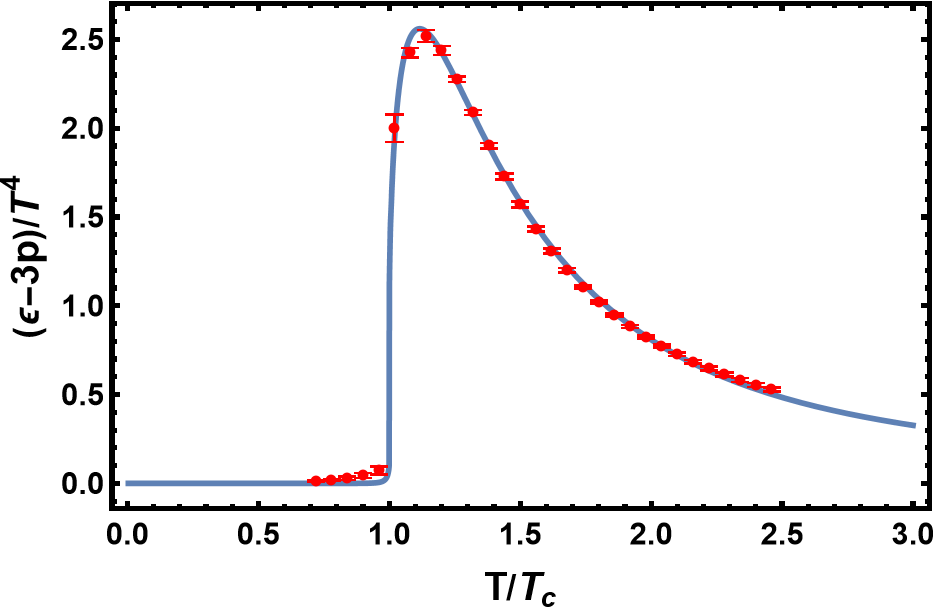}
  \caption{Pure-glue calibration of the EMD sector at $\mu=0$: comparison between the holographic EoS and the lattice Yang--Mills EoS \cite{Caselle:2018kap}. Upper left panel: the ratio of entropy density over cubic temperature, $s/T^3$. Upper right panel: the ratio of pressure over quartic temperature, $p/T^4$. Lower left panel: the ratio of internal energy density over quartic temperature, $\epsilon/T^4$. Lower right panel: the trace anomaly over quartic temperature, $I/T^4=(\epsilon-3p)/T^4$. In all panels the horizontal axis is the scaled temperature $T/T_c$. The blue solid lines are our holographic results. The red points (with error bars) are the lattice data from Ref.~\cite{Caselle:2018kap}.}
  \label{fig_pureglue_EoS_vs_lattice}
\end{figure*}

With the best-fit parameter set, we compute the EoS at $\mu=0$ and compare it with the lattice Yang--Mills result, as shown in Fig.~\ref{fig_pureglue_EoS_vs_lattice}. The red points are the lattice $SU(3)$ data from Ref.~\cite{Caselle:2018kap}.

\section{Coupled \texorpdfstring{EMD$+$KKSS}{EMD+KKSS} system with back-reaction and the \texorpdfstring{$N_f=2$}{Nf=2} EoS}
\label{sec_back-reaction_nf2}

\begin{sloppypar}
In this section, we go beyond the probe approximation and solve the coupled EMD$+$KKSS equations with back-reaction.
The logic is the following:
\end{sloppypar}
\begin{itemize}
  \item we first fix the EMD sector by fitting the pure-glue lattice EoS at $\mu=0$, as shown in Sec.~\ref{sec_pureglue};
  \item then we turn on the KKSS sector and solve the coupled equations, and interpret the back-reacted solution as a dual description of the $N_f=2$ theory.
\end{itemize}

\subsection{Coupled equations and numerical setup}
\label{subsec_back-reaction_EoMs}

\begin{sloppypar}
From the numerical point of view, turning on the flavor back-reaction turns the thermodynamic problem into a genuinely coupled boundary-value system: one has to solve five second-order ODEs for $\{A_E(z),f(z),\phi(z),A_t(z),\chi(z)\}$ subject to UV AdS normalizations and horizon regularity conditions, and the solution determines $(T,n_B,\sigma_u)$ simultaneously.
This coupled structure (together with the fractional-power UV asymptotics induced by $\Delta_\phi=3.8$) makes the back-reacted step substantially more constrained and more delicate than the probe calculation, and it also provides a clean arena to diagnose which model ingredients are still missing.
\end{sloppypar}

Varying the total action in Eq.~(\ref{action_total}) gives a coupled system of equations for $\{A_E(z),f(z),\phi(z),A_t(z),\chi(z)\}$.
In the present back-reaction study we focus on the thermodynamic sector and keep the dominant $\chi$ contribution in the KKSS action, while the meson and baryon excitations are neglected, as explained in Sec.~\ref{subsec_flavor_part}.
The $\chi$ equation of motion is given in Eq.~(\ref{EoMs_back_6}), and the back-reaction enters through the modified Einstein equations, where the energy-momentum tensor receives additional contributions from the flavor sector.

\begin{sloppypar}
For completeness, we keep the general form of the coupled equations below.
However, in the EoS comparison reported in this section we work at $\mu=0$ and finite temperature, thus $A_t(z)\equiv 0$ and the Maxwell sector decouples.
\end{sloppypar}
The numerical integration is performed by imposing UV regularity and horizon regularity, and one of the Einstein equations is used as a constraint to check the solutions.

More explicitly, we adopt the standard black hole ansatz in the Einstein frame, Eq.~(\ref{metric_Einstein}), with horizon located at $z=z_h$ where $f(z_h)=0$. The Hawking temperature is obtained from Eq.~(\ref{temperature}). At the UV boundary $z\to 0$, we impose the asymptotic AdS behavior and the UV expansions summarized in Eq.~(\ref{eq_UV_asymptotic_expansions}). At the horizon, all fields are required to be regular, which fixes the near-horizon expansions and reduces the number of free shooting parameters.

It is noticed that the coupled system contains redundant equations due to diffeomorphism invariance. In practice, we solve a convenient subset of the coupled equations and use one Einstein equation as a constraint (analogous to the role of Eq.~(\ref{EoMs_2_3}) in the probe EMD system) to monitor the numerical accuracy of the solution.

\subsection{KKSS input: \texorpdfstring{$\beta(\Phi)$}{beta(Phi)} and \texorpdfstring{$V_X$}{VX}}
\label{subsec_back-reaction_KKSS_input}

The KKSS sector is specified by the function $\beta(\Phi)$ and the scalar potential $V_X^s(X;\Phi,F^2)$ (or equivalently its Einstein-frame form $V_X^E$).
In the $N_f=2$ case, the bulk scalar $X$ is a $2\times 2$ matrix. In the thermodynamic sector considered here we take its vacuum expectation value in the diagonal form $\langle X\rangle=\frac{\chi(z)}{2}I_2$, thus only $\Tr V_X$ enters.
In the coupled $N_f=2$ calculation reported here, we fix the KKSS polynomial potential parameters $\lambda_2=-3$ and $\lambda_4=168/5$ (see Eq.~(\ref{potent_X}) and the discussion in Sec.~\ref{subsec_flavor_part}), and set the covariant-derivative coupling $g_c=0$.
For the flavor-backreaction strength we adopt the smooth ansatz in Eq.~(\ref{beta_ansatz}) with $\beta_2=1$ and $\beta_3=50$, while $\beta_1$ is varied to perform a comparison study:
\begin{align}
  \beta_1 &\in \left\{21,\,23,\,25\right\},
  \nonumber\\
  \beta_2 &= 1,
  \nonumber\\
  \beta_3 &= 50,
  \nonumber\\
  g_c &= 0,
  \nonumber\\
  \lambda_4 &= \frac{168}{5},
  \nonumber\\
  m_u&=24.826~\mathrm{MeV},
  \label{eq_KKSS_params_back-reacted}
\end{align}
where $\beta_1$ controls the overall magnitude of $\beta(\Phi)$ and thus the strength of the flavor back-reaction; the three values above are used to assess the corresponding systematic uncertainty in the $N_f=2$ EoS comparison.

\begin{sloppypar}
Ref.~\cite{Burger:2014xga} reported the $N_f=2$ lattice EoS for three sets of bare parameters (corresponding to different pion masses), and in this work we compare with their $m_\pi=360~\mathrm{MeV}$ ensemble.
Accordingly, the input light-quark mass in our holographic calculation is not taken at the physical point.
Since the present thermodynamic truncation does not compute the pion mass directly, we fix the degenerate $u/d$ current quark mass $m_u$ using the Gell-Mann--Oakes--Renner relation, i.e. $m_u\propto m_\pi^2$ at fixed $f_\pi$ and condensate normalization, and rescale from the physical pion mass.
This gives $m_u=24.826~\mathrm{MeV}$, which is used as the source parameter $c_{\chi,5}$ in the coupled $N_f=2$ study reported below.
\end{sloppypar}

\subsection{Two-flavor EoS and comparison with lattice}
\label{subsec_back-reaction_results}

\begin{sloppypar}
With the EMD parameters fixed by the pure-glue calibration, we solve the coupled system and compute the EoS at $\mu=0$.
We then compare the result with the lattice EoS of $N_f=2$ QCD. We find that a visible mismatch remains: by manually tuning the KKSS parameters one can reproduce the overall trend, but quantitative differences persist in some temperature ranges.
We also tried a hybrid optimization strategy for the KKSS parameters, but the improvement is limited and the fit quality is still not satisfactory.
This mismatch may indicate limitations of the current ansatz for $\beta(\Phi)$ and/or $V_X$ in the present truncation of the flavor sector.
In particular, as will be seen from the high-temperature behavior below, the coupled results tend to a nearly $\beta_1$-insensitive plateau close to the $\beta_1=0$ (pure-glue) curve, suggesting that the present KKSS truncation may not provide sufficient high-$T$ flavor contribution in the thermodynamics.
\end{sloppypar}

\subsubsection{Sensitivity to the back-reaction strength parameter \texorpdfstring{$\beta_1$}{beta1}}
\label{subsubsec_beta1_sensitivity}

\begin{figure}[tbp]
  \centering
  \includegraphics[width=0.84\linewidth,clip=true,keepaspectratio=true]{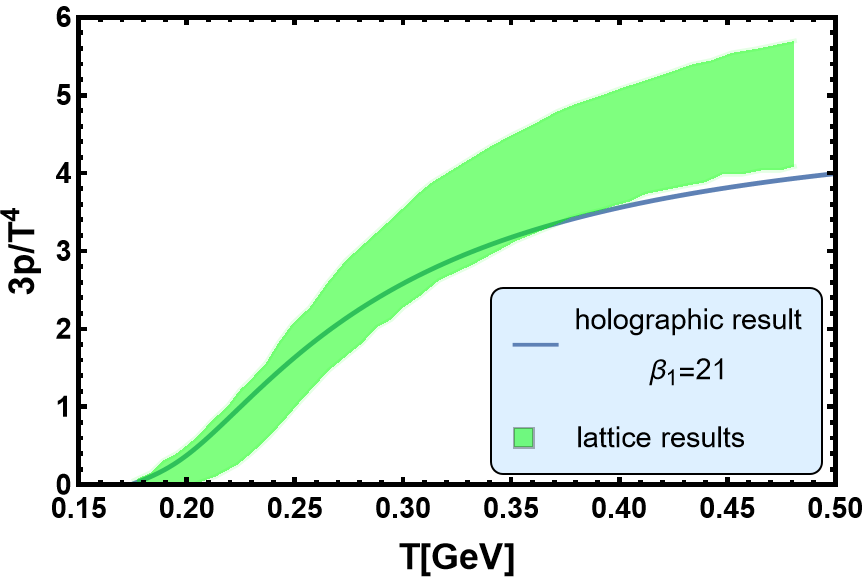}
  \vspace{0.08cm}\\
  \includegraphics[width=0.84\linewidth,clip=true,keepaspectratio=true]{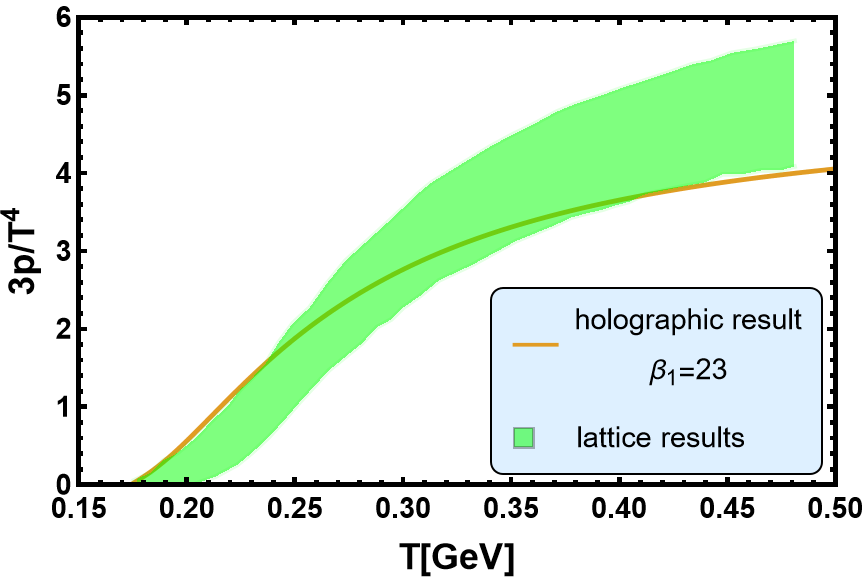}
  \vspace{0.08cm}\\
  \includegraphics[width=0.84\linewidth,clip=true,keepaspectratio=true]{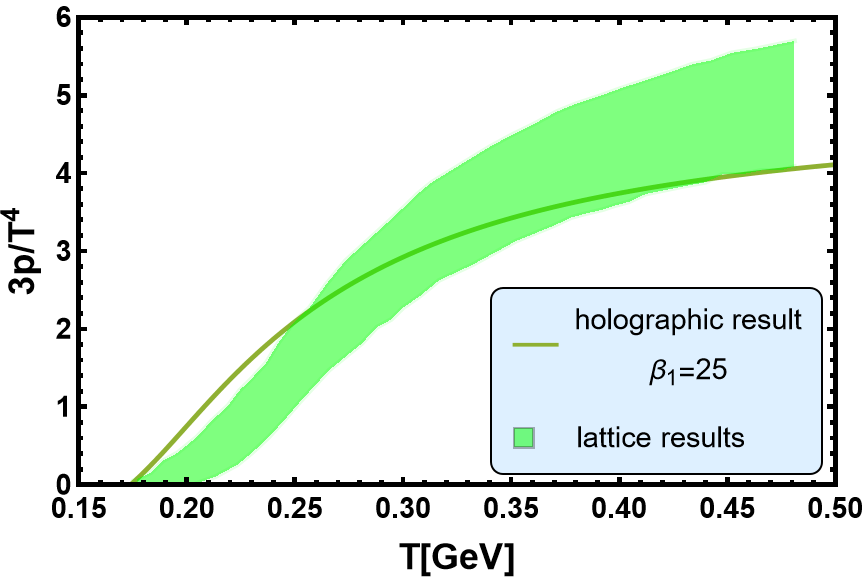}
  \caption{Coupled EMD$+$KKSS system with back-reaction at $\mu=0$:
  comparison of $3p/T^4$ with the $N_f=2$ lattice QCD result of Ref.~\cite{Burger:2014xga} (green band) for three values of the flavor back-reaction strength.
  Upper: $\beta_1=21$; middle: $\beta_1=23$; lower: $\beta_1=25$.
  The solid line in each panel is the corresponding holographic result.}
  \label{fig_beta1_pressure_vs_lattice}
\end{figure}

In Fig.~\ref{fig_beta1_pressure_vs_lattice}, we compare the back-reacted holographic result for $3p/T^4$ with the $N_f=2$ lattice band of Ref.~\cite{Burger:2014xga},
for three representative values $\beta_1=21,23,25$.
We find that the holographic EoS captures the overall trend of the lattice result, but it is not completely inside the lattice uncertainty band in the full temperature window.
More concretely, increasing $\beta_1$ raises the holographic curve. This improves the agreement at higher temperatures, but it can also spoil the agreement at lower temperatures:
for $\beta_1=21$ the low-$T$ region is closer to the lattice band while the high-$T$ region tends to undershoot, whereas for $\beta_1=25$ the curve is lifted and becomes closer to the band at higher $T$
at the price of overshooting at lower $T$.

The origin of this behavior can be traced back to the role of $\beta_1$ in the KKSS sector.
The function $\beta(\Phi)$ in Eq.~(\ref{action_matter}) controls the coupling strength between the flavor sector and the EMD background.
With the smooth ansatz in Eq.~(\ref{beta_ansatz}), $\beta(\Phi)$ interpolates from $0$ in the UV to $\beta_1$ in the IR, so $\beta_1$ sets the overall magnitude of the flavor back-reaction
entering the coupled Einstein equations, e.g. the source term $\propto \beta(\Phi)e^{\Phi}T_{MN}^{\chi}$ in Eq.~(\ref{EoMs_back_tensor_Einstein}).
In this sense, $\beta_1$ is a phenomenological knob for the back-reaction strength (rather than a QCD beta-function coefficient).
It does not act as a trivial multiplicative factor on the EoS: even at $\mu=0$ it changes the full coupled solution and hence modifies the horizon data and the map between $T$ and $z_h$,
leading to a temperature-dependent deformation of dimensionless ratios such as $p/T^4$ and $\epsilon/T^4$.
Since the horizon probes different dilaton values at different temperatures, the effective back-reaction strength is $T$ dependent, which naturally explains why a single parameter $\beta_1$
cannot uniformly fix the curve in all temperature regions within the present ansatz.

\begin{figure}[tbp]
  \centering
  \includegraphics[width=0.84\linewidth,clip=true,keepaspectratio=true]{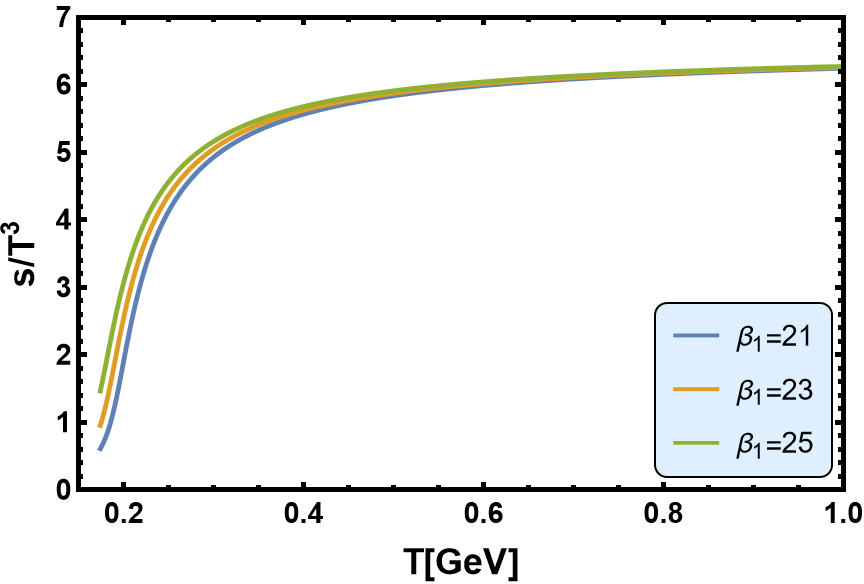}
  \vspace{0.08cm}\\
  \includegraphics[width=0.84\linewidth,clip=true,keepaspectratio=true]{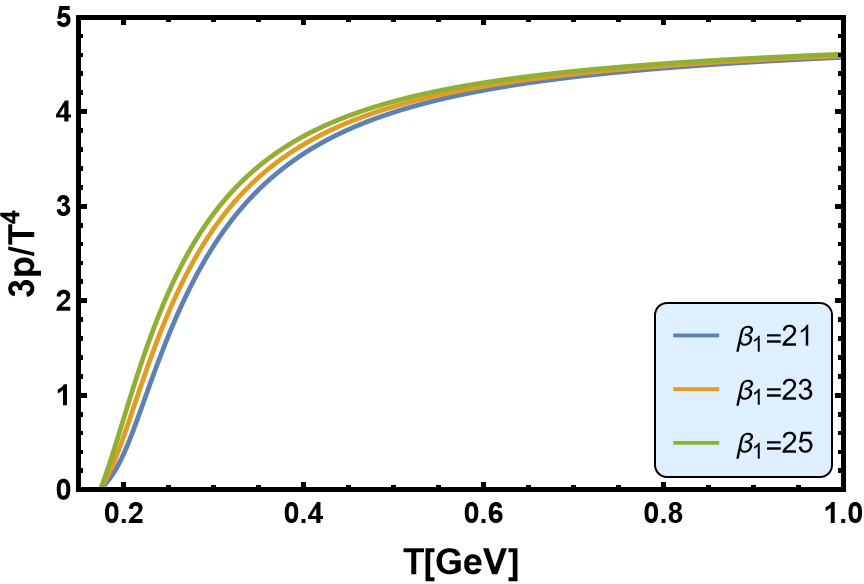}
  \vspace{0.08cm}\\
  \includegraphics[width=0.84\linewidth,clip=true,keepaspectratio=true]{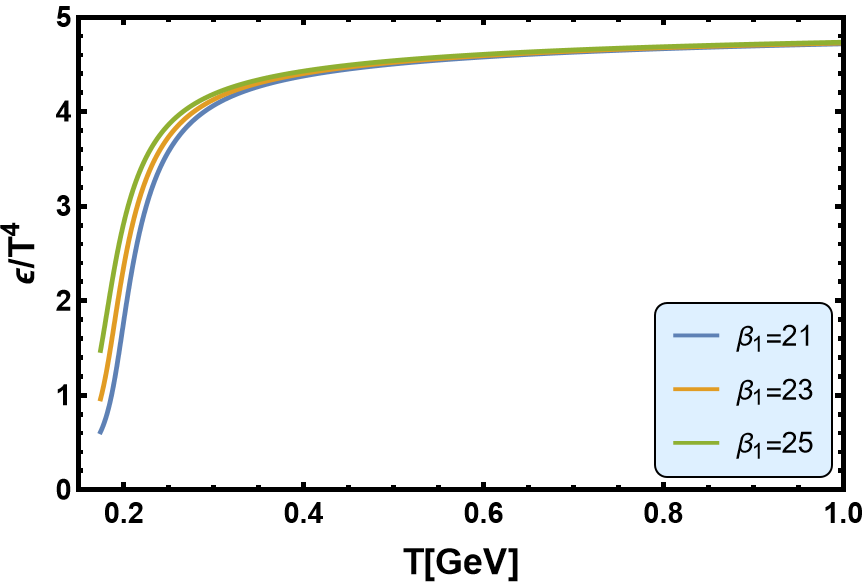}
  \vspace{0.08cm}\\
  \includegraphics[width=0.84\linewidth,clip=true,keepaspectratio=true]{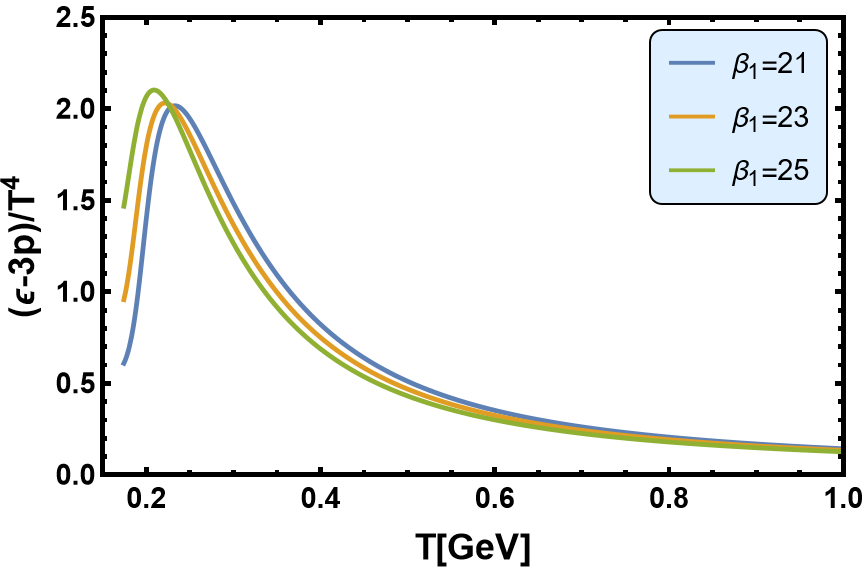}
  \caption{Dependence on the flavor back-reaction parameter $\beta_1$ in the back-reacted $N_f=2$ study at $\mu=0$.
  From top to bottom: $s/T^3$, $3p/T^4$, $\epsilon/T^4$, and $(\epsilon-3p)/T^4$.
  In each panel, the three curves correspond to $\beta_1=21,23,25$, and the horizontal axis is $T$ in units of GeV.}
  \label{fig_beta1_multi_observables}
\end{figure}

To further clarify the impact of $\beta_1$, in Fig.~\ref{fig_beta1_multi_observables} we plot $s/T^3$, $3p/T^4$, $\epsilon/T^4$, and $(\epsilon-3p)/T^4$ for $\beta_1=21,23,25$.
We see a monotonic increase of $s/T^3$, $3p/T^4$ and $\epsilon/T^4$ with $\beta_1$ in the whole temperature range shown.
The interaction measure $(\epsilon-3p)/T^4$ is more sensitive around the crossover: increasing $\beta_1$ changes the peak region more visibly than the high-$T$ tail.
Therefore, tuning only the overall IR magnitude $\beta_1$ tends to shift the EoS upward in a correlated way across observables, but it does not provide enough freedom to
simultaneously match the low-$T$ and high-$T$ parts of the lattice EoS.
This suggests that a better quantitative agreement in the back-reacted $N_f=2$ setup likely requires additional flexibility,
e.g. varying the $\Phi$ dependence of $\beta(\Phi)$ (parameters $\beta_2,\beta_3$) and/or enlarging the ansatz for $V_X$.

\begin{figure}[tbp]
  \centering
  \includegraphics[width=0.82\linewidth,clip=true,keepaspectratio=true]{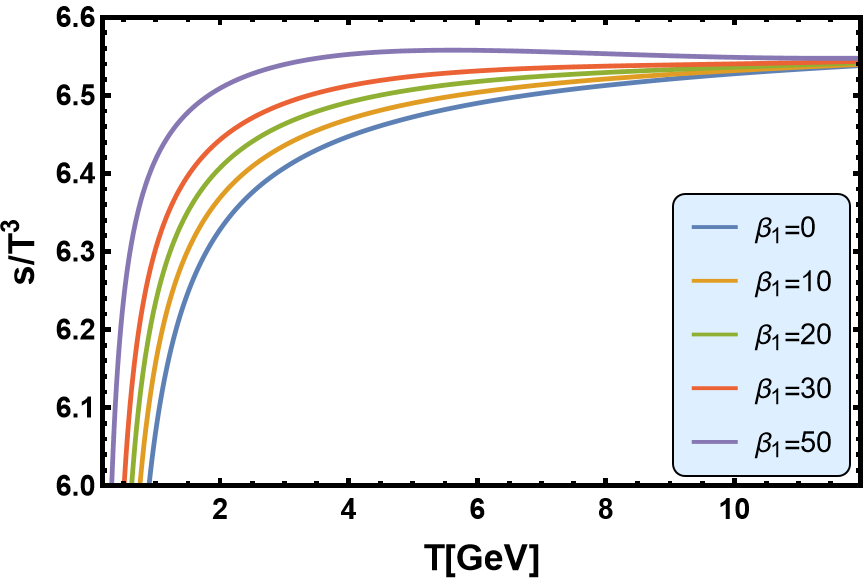}
  \vspace{0.07cm}\\
  \includegraphics[width=0.82\linewidth,clip=true,keepaspectratio=true]{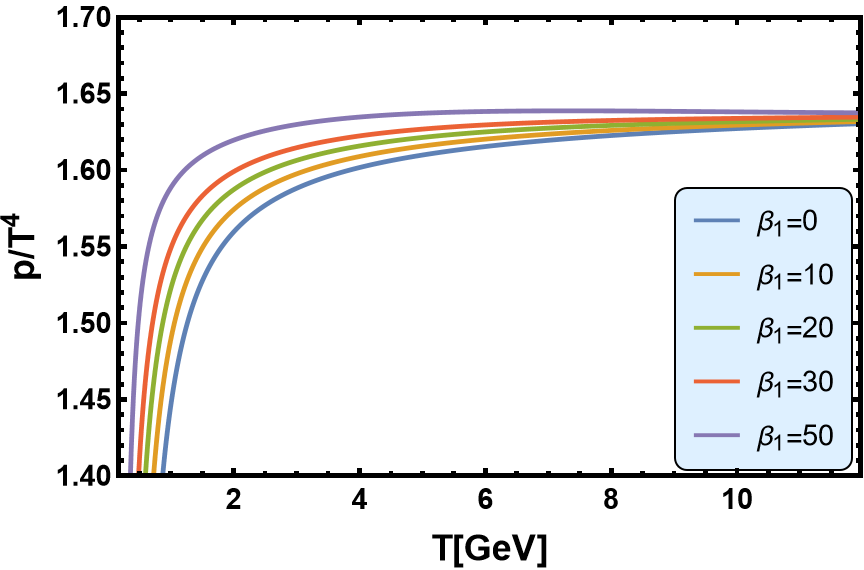}
  \vspace{0.07cm}\\
  \includegraphics[width=0.82\linewidth,clip=true,keepaspectratio=true]{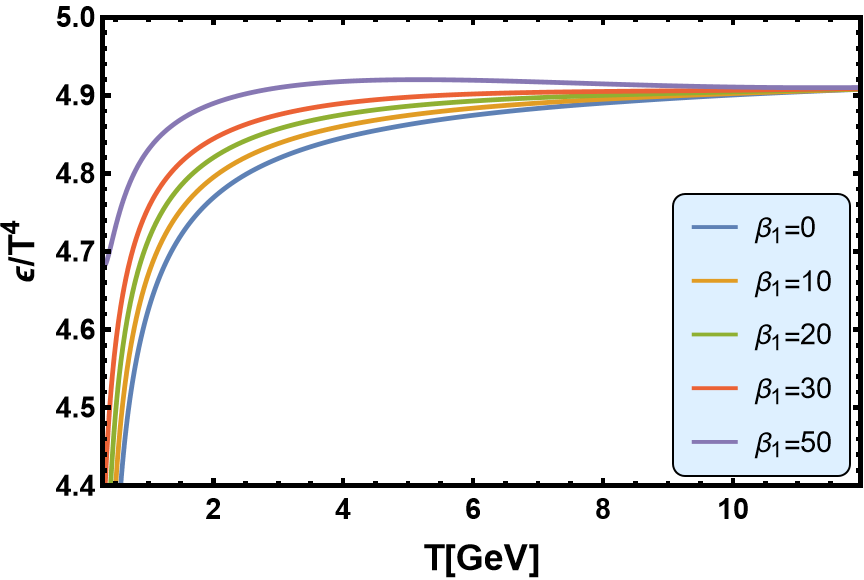}
  \vspace{0.07cm}\\
  \includegraphics[width=0.82\linewidth,clip=true,keepaspectratio=true]{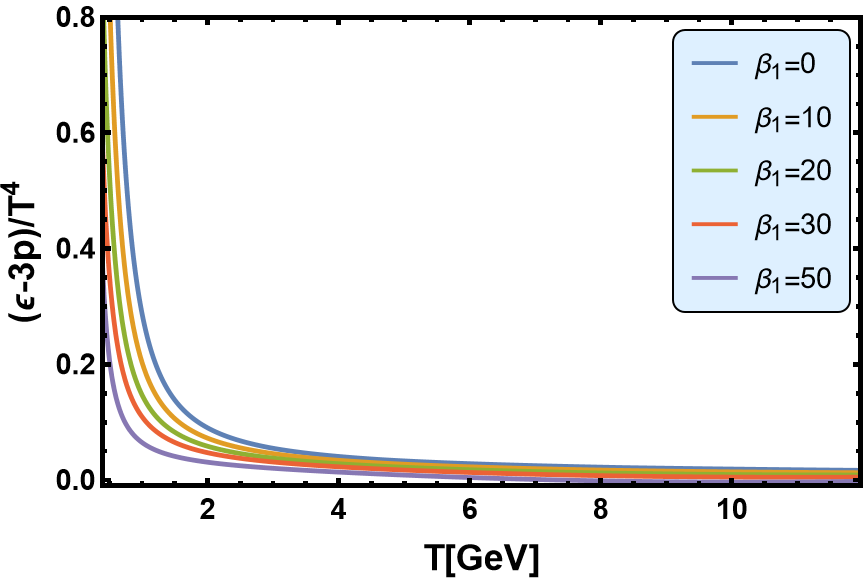}
  \caption{High-temperature behavior in the back-reacted $N_f=2$ study at $\mu=0$ for different values of the flavor back-reaction strength parameter $\beta_1$.
  From top to bottom: $s/T^3$, $p/T^4$, $\epsilon/T^4$, and $(\epsilon-3p)/T^4$.
  In each panel, the five curves correspond to $\beta_1=0,10,20,30,50$, and the horizontal axis is $T$ in units of GeV.
  It is noticed that the curve at $\beta_1=0$ corresponds to switching off the flavor coupling, $\beta(\Phi)=0$, thus it reduces to the pure EMD (pure-glue, $N_f=0$) system.}
  \label{fig_beta1_highT_observables}
\end{figure}

\begin{sloppypar}
It is also instructive to examine the high-temperature behavior.
As shown in Fig.~\ref{fig_beta1_highT_observables}, for sufficiently large $T$ the curves for different values of $\beta_1$ approach nearly the same constant (in particular, they approach the $\beta_1=0$ curve), while the interaction measure $(\epsilon-3p)/T^4$ rapidly decays to zero.
It is noticed that $\beta_1=0$ corresponds to switching off the flavor coupling, $\beta(\Phi)=0$, thus it reduces to the pure EMD (pure-glue, $N_f=0$) system.
This indicates that, within the present truncation of the flavor sector to the $\chi$ contribution, the effect of the back-reaction parameter $\beta_1$ becomes weak in the high-$T$ region.
\end{sloppypar}

\begin{sloppypar}
This behavior can be understood from the structure of the KKSS action adopted here.
In our setup, the flavor sector enters the thermodynamics through $\beta(\Phi)$ and the $\chi$ sector, and the polynomial potential $V_X$ in Eq.~(\ref{potent_X}) does not contain a $\chi$-independent constant term.
As a result, in the high-$T$ deconfined region where $\chi$ is small and the horizon moves toward the UV, the flavor contribution to the stress tensor (and hence to $p/T^4$, $s/T^3$, and $\epsilon/T^4$) is suppressed, so the asymptotic plateaus are dominated by the EMD sector that has been fixed by the pure-glue calibration.
This is qualitatively different from the Stefan--Boltzmann (SB) limit of massless QCD, where the pressure receives an ${\cal O}(N_c N_f)$ contribution from quark degrees of freedom.
For reference, for ${\rm SU}(N_c)$ with $N_f$ massless quark flavors one has
\begin{align}
  \left.\frac{s}{T^3}\right|_{\mathrm{SB}}
  &=4\left.\frac{p}{T^4}\right|_{\mathrm{SB}},
  \nonumber\\
  \left.\frac{p}{T^4}\right|_{\mathrm{SB}}
  &=\frac{\pi^2}{45}(N_c^2-1)+\frac{7\pi^2}{180}N_c N_f,
  \nonumber\\
  \left.\frac{\epsilon}{T^4}\right|_{\mathrm{SB}}
  &=3\left.\frac{p}{T^4}\right|_{\mathrm{SB}},
  \nonumber\\
  \left.\frac{\epsilon-3p}{T^4}\right|_{\mathrm{SB}}
  &=0,
  \label{eq_SB_limit}
\end{align}
Therefore, the fact that our back-reacted curves tend to a $\beta_1$-insensitive plateau in the high-$T$ region provides a simple diagnostic for the present model ansatz: to reproduce the correct high-$T$ flavor contribution, it may be necessary to enlarge the flavor action, e.g. by introducing an additional $\chi$-independent term in the flavor potential (which can be motivated from brane/tension terms in DBI-inspired constructions) and/or by including additional flavor-sector degrees of freedom beyond the $\chi$ truncation.
\end{sloppypar}

\subsection{Parameter summary}
\label{subsec_parameter_summary}

For convenience, we summarize the parameter inputs used in different steps.
\begin{table*}[htbp]
  \caption{Summary of parameter choices used in different steps. In the probe approximation, $V_{\phi}(\phi)$ and $h_{\phi}(\phi)$ are obtained as discrete samples from the neural ODE calibration and then parametrized. In the pure-glue step, the EMD parameters are fixed by a hybrid optimization. In the back-reacted $N_f=2$ step, the EMD sector is kept fixed and the KKSS couplings (including the back-reaction strength encoded in $\beta(\Phi)$) are varied for comparison with the lattice EoS.}
  \label{tab_param_summary}
  \begin{ruledtabular}
    \begin{tabular}{l p{0.72\textwidth}}
      Sector / step & Key inputs \\
      \hline
      Probe $(2+1)$, finite $T,\mu$
      & $V_{\phi}(\phi)$ and $h_{\phi}(\phi)$ from neural-ODE calibration (Sec.~\ref{subsec_neuralODE}); see also Eqs.~(\ref{eq_probe_VG_param})--(\ref{eq_probe_bestfit_constants}) \\
      Pure glue, $\mu=0$
      & $V_{\phi}(\phi)$ ansatz in Eq.~(\ref{eq_pureglue_VG_ansatz}) with $\Delta_{\phi}=3.8$; best-fit normalizations $(\Lambda,G_5)$ in Eq.~(\ref{eq_pureglue_norm_params}); best-fit potential parameters $(v_0,a_4,a_6,a_8,a_{10})$ in Eq.~(\ref{eq_pureglue_V_params}) \\
      Back-reacted $N_f=2$, $\mu=0$
      & $\beta(\Phi)$ in Eq.~(\ref{beta_ansatz});
      $g_c$ in Eq.~(\ref{covar_der});
      $(\lambda_2,\lambda_4)$ in Eq.~(\ref{potent_X});
      $m_u=24.826~\mathrm{MeV}$ (fixed by the $m_\pi=360~\mathrm{MeV}$ ensemble of Ref.~\cite{Burger:2014xga} via the GOR relation);
      and the parameter values used in this work are summarized in Eq.~(\ref{eq_KKSS_params_back-reacted}) \\
    \end{tabular}
  \end{ruledtabular}
\end{table*}
\section{Conclusion and outlook}
\label{sec_conclusion}

\begin{sloppypar}
In this work, we studied the QCD equation of state in a bottom-up holographic framework built from an EMD sector and an improved KKSS flavor sector.
The analysis consists of three steps based on the same setup but different approximations.
In the probe approximation, we calibrated the EMD model to the lattice EoS of $(2+1)$-flavor QCD at finite temperature and finite baryon chemical potential by using a neural ODE strategy, and obtained good agreement in the calibrated domain.
Treating the EMD system as an effective dual description of pure Yang--Mills theory, we then fixed the EMD parameters by fitting the $\mu_B=0$ lattice pure-glue EoS with a hybrid optimization method.
Finally, we went beyond the probe approximation and solved the coupled EMD$+$KKSS equations with back-reaction at $\mu_B=0$; with the EMD sector fixed by the pure-glue step, we tuned the KKSS parameters to the lattice EoS of two-flavor QCD, where a visible mismatch remains.
\end{sloppypar}

\begin{sloppypar}
It is noticed that the present mismatch is not only a matter of tuning.
With the EMD sector fixed by the pure-glue calibration, the coupled results at high temperature tend to approach a nearly $\beta_1$-insensitive plateau close to the $\beta_1=0$ (pure-glue) baseline, while $(\epsilon-3p)/T^4$ rapidly decays toward zero.
This provides a simple diagnostic for the present coupled ansatz and indicates that, within the present truncation of the flavor sector to the $\chi$ contribution and the polynomial potential in Eq.~(\ref{potent_X}), the high-$T$ flavor contribution in the thermodynamics is suppressed compared with the expectation from QCD.
Therefore, the mismatch may have several origins, such as limitations of the current ansatz for \texorpdfstring{$\beta(\Phi)$}{beta(Phi)} and \texorpdfstring{$V_X$}{VX}, parameter degeneracy, and missing operators in the bottom-up construction.
Possible next moves include:
\end{sloppypar}

\begin{sloppypar}
In particular, the nearly flavor-blind high-$T$ plateau may indicate that the present KKSS-type truncation lacks a \emph{$\chi$-independent} contribution to the flavor free energy.
A natural extension, motivated by the generic structure of DBI/tachyon effective actions (where the tachyon potential typically satisfies $V_f(\chi=0)=1$), is to allow a brane-tension-like term in the Einstein-frame flavor potential,
\begin{align}
  V_X^E(\chi,\Phi,F_E^2)=V_0^E(\Phi)+\lambda_2|\chi|^2+\lambda_4|\chi|^4+\cdots,
\end{align}
so that the flavor sector can contribute to the thermodynamics even in the chirally symmetric background ($\chi=0$), providing an adjustable \texorpdfstring{$\mathcal{O}(N_cN_f)$}{O(Nc Nf)} component in the pressure/entropy at high temperature.
It is worth noting that strongly coupled holography is not expected to coincide with the Stefan--Boltzmann limit, but the high-$T$ behavior does provide a clean diagnostic and constrains admissible flavor-sector completions.
\end{sloppypar}

\begin{sloppypar}
Another closely related issue is the extraction of chiral observables in the present back-reacted setup.
Since we adopt a non-integer UV scaling dimension $\Delta_{\phi}=3.8$, the near-boundary expansions of the background fields involve fractional powers in $z$.
Through the coupled equations, these fractional-power terms propagate into the UV expansion of the chiral scalar $\chi(z)$ and spoil a clean separation of the integer-power modes.
As a result, a naive identification of the chiral condensate with the coefficient of the $z^3$ term becomes numerically ill-conditioned.
A systematic treatment should instead define the condensate from the renormalized canonical momentum (equivalently, from the renormalized on-shell action) by introducing appropriate counterterms at a UV cutoff $z=\epsilon$ and taking $\epsilon\to 0$.
We leave such a holographic-renormalization analysis, as well as the associated chiral transition study, to future work.
\end{sloppypar}
\begin{itemize}
  \item enlarging the functional space of $\beta(\Phi)$ and/or $V_X$ (allowing more terms and/or different IR behaviors);
  \item allowing for a $\chi$-independent term $V_0^E(\Phi)$ in $V_X^E$ (brane-tension contribution) to tune the high-$T$ flavor plateau;
  \item adding additional observables to constrain the flavor sector, e.g. chiral condensate and chiral transition;
  \item performing holographic renormalization for the back-reacted flavor sector to define $\sigma_u$ from the renormalized canonical momentum in the presence of fractional-power UV asymptotics;
  \item including susceptibilities and/or more data at small $\mu$ in the fitting strategy;
  \item extending the back-reacted calculation to finite baryon chemical potential.
\end{itemize}


\begin{acknowledgments}
  We thank Maria Paola Lombardo for helpful discussions.
\end{acknowledgments}


\bibliographystyle{unsrtnat}
\bibliography{references}

\end{document}